\documentclass[aps,prb,twocolumn,showpacs,superscriptaddress]{revtex4-1}

\usepackage{amsmath} \usepackage{graphicx} \usepackage{hyperref}
\usepackage{color} \usepackage{algorithm} \usepackage{algorithmic}
\usepackage{amsfonts} \usepackage{amssymb} \usepackage[caption=false]{subfig}
\usepackage{bbold}

\newcommand{\dcaplus}{DCA$^+$\:} 

\begin{document}

\title{\dcaplus: Dynamical Cluster Approximation with continuous lattice
self-energy.} \author{Peter Staar} \affiliation{Institute for Theoretical
Physics, ETH Zurich, 8093 Zurich, Switzerland} \author{Thomas Maier}
\affiliation{Computer Science and Mathematics Division, Oak Ridge National
Laboratory, Oak Ridge TN, 37831, USA} \affiliation{Center for Nanophase
Materials Sciences, Oak Ridge National Laboratory, Oak Ridge TN, 37831, USA}
\author{Thomas C. Schulthess} \affiliation{Institute for Theoretical Physics, ETH
Zurich, 8093 Zurich, Switzerland} \affiliation{Computer Science and
Mathematics Division, Oak Ridge National Laboratory, Oak Ridge TN, 37831, USA}
\affiliation{Swiss National Supercomputing
Center, ETH Zurich, 6900 Lugano, Switzerland} 
\date{\today }

\begin{abstract} The dynamical cluster approximation (DCA) is a systematic
extension beyond the single site approximation in dynamical mean field theory
(DMFT), to include spatially non-local correlations in quantum many-body
simulations of strongly correlated systems. We extend the DCA with a continuous
lattice self-energy in oder to achieve better convergence with cluster size. The
new method, which we call \dcaplus, cures the cluster shape dependence problems
of the DCA, without suffering from causality violations of previous attempts to
interpolate the cluster self-energy. A practical approach based on standard
inference techniques is given to deduce the continuous lattice self-energy from
an interpolated cluster self-energy. We study the pseudogap region of a
hole-doped two-dimensional Hubbard model and find that in the \dcaplus
algorithm, the self-energy and pseudo-gap temperature $T^*$ converge
monotonously with cluster size. Introduction of a continuous lattice self-energy
eliminates artificial long-rage correlations and thus significantly reduces the
sign problem of the quantum Monte Carlo cluster solver in the \dcaplus algorithm
compared to the normal DCA. Simulations with much larger cluster sizes thus
become feasible, which, along with the improved convergence in cluster size,
raises hope that precise extrapolations to the exact infinite cluster size limit
can be reached for other physical quantities as well. \end{abstract}


\maketitle

\section*{Introduction:}

The study of interacting electrons in a crystalline solid remains one of the
most challenging problems of condensed matter physics. On a purely theoretical
level, these models give us insight on spontaneous symmetry breaking, which
leads to new ground states with exciting properties such as superconductivity.
On a more practical level, the lattice-models allow us to better understand
materials in which the correlations between electrons determines its physical
properties. The most famous examples of such materials are the high-$T_c$
cuprates\cite{Bednorz1986} and the recently discovered
pnictides\cite{Ozawa2008}. A better understanding of how the Cooper pairs are
formed in these materials might lead us in the future to the creation of new
materials with higher superconducting transition temperature.

One of the methods of choice to investigate interacting electrons on a
lattice-model is the DMFT\cite{Georges1996RMP}, which, in conjunction with model
parameters derived from first principles electronic structure
calculations\cite{Anisimov1997,Anisimov1999,Kotliar2004,Kotliar2006,
Lichtenstein1998PRB}, is now capable of predicting spectral properties of
transition metal oxides \cite{Kunes2008} and heavy fermion
materials\cite{Dai2003Science,Held2001PRL,Held2001PRL2,McQueeneyPRL,
Savrasov2001Nat,Shim2007Nat}. The study of the paring mechanism in
superconductors, however, requires inclusions of dynamic correlations between
lattice sites, and hence the extension of DMFT beyond the single site
approximation. To this end, several quantum cluster extensions to DMFT have been
developed during the past fifteen years
\cite{Hettler1998PRB,Kotliar2001,Lichtenstein2000PRB,Maier2005RMP}. Among these
is the DCA \cite{Hettler2000PRB, Jarrell2001PRB, Maier2005RMP}, a systematic
extension to DMFT that includes non-local correlations through coarse-graining
in momentum space. The DCA relies on the assumption that the self-energy
function is a localized function in real space. In infinite dimensions, it has
been proven that the self-energy $\Sigma$ is a delta-function in real space
\cite{Metzner1989PRL}, in which case this assumption trivially holds. In
practice, we see that the locality increases with increasing dimension. This
explains why a single-site DMFT approach generally works better for 3D
materials, but fails to describe materials of quasi 1D or 2D nature.

\begin{figure}[t] \begin{center}
\includegraphics[width=0.5\textwidth]{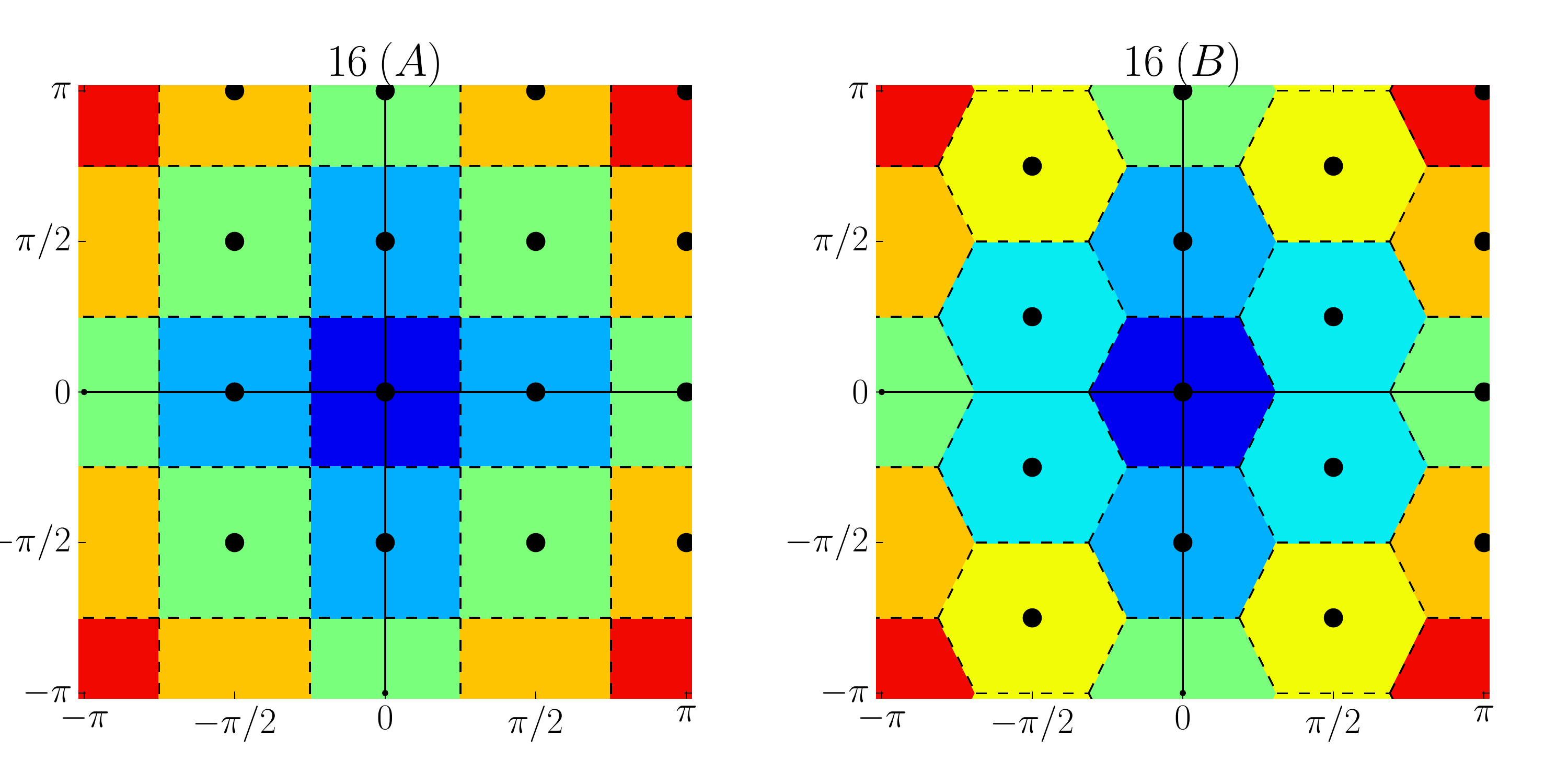} 
\end{center}
\caption{\label{fig:clusters} The positions of the cluster momenta $\{\vec{K}\}$
and shape of the patches for two 16 site DCA-clusters. Notice that the 16B site
cluster does not have the same point-group symmetry as the Brillouin-zone,
leading to a lattice self-energy with a lower symmetry.} \end{figure}

The Dynamical Cluster Approximation was developed to study materials of 2D
nature, by allowing the self-energy to be non-local. In the DCA, the infinite
lattice-problem is reduced to a finite size quantum cluster impurity with
periodic boundary conditions, embedded into a self-consistent mean-field. This
reduction is achieved via a coarse-graining procedure of the Green's function,
in which the Brillouin zone is divided into $N_c$ patches and the self-energy
$\Sigma$ is assumed to be constant on these patches. In this way, all
correlations within the cluster are dealt with exactly, while long-range
correlations outside the cluster are described via a mean-field. If the cluster
impurity problem is solved exactly, such as with  Quantum Monte Carlo (QMC)
integration, the DCA will reproduce the exact solution of the lattice model in
the limit of infinite cluster size.

\begin{figure*}[t] 
\begin{center}
\includegraphics[width=\textwidth]{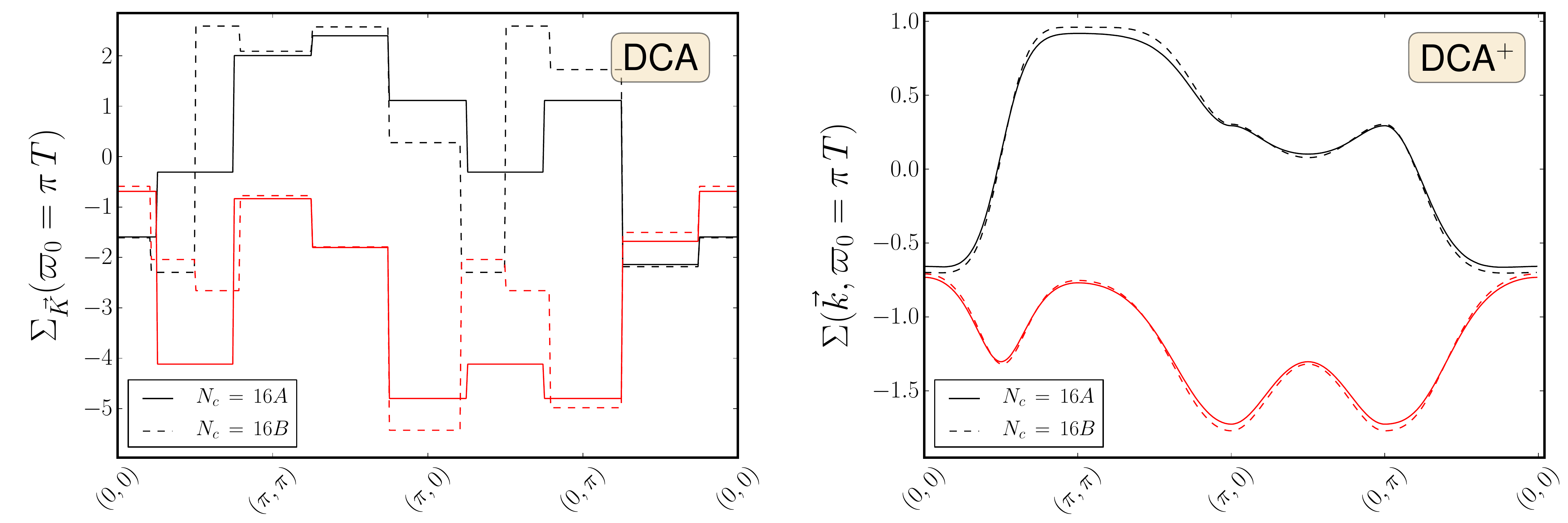} 
\end{center}
\caption{\label{fig:sigma_band_structure_DCA} Momentum-dependence of the DCA and
\dcaplus self-energies (red and black represent the imaginary and real part) 
calculated on the $16A$ and $16B$ clusters in a
half-filled Hubbard model with nearest neighbor hopping $t=1$,  Coulomb
interaction $U/t=7$ and next-nearest neighbor hopping $t'/t=-0.15$ at a
temperature $T=0.2$. For the DCA, one clearly sees a large difference between
the self-energies of the two clusters at the section $(\pi,0) \rightarrow
(0,\pi)$, which is close to the Fermi-surface and thus physically the most
relevant part of the self-energy. In the \dcaplus, the self-energies of the two
clusters agree very well.} \end{figure*}

In practice, the fermionic sign problem\cite{Loh1990PRB,Troyer:2005ui} imposes
an upper-bound to the cluster-size and a lower bound to the temperature which
can be accessed. While small clusters have proven to give us an excellent
qualitative insight on the physical phenomena \cite{Maier2005RMP}, most physical
quantities, such as the superconducting transition temperature $T_c$, converge
poorly on the available small clusters \cite{Maier2005PRL}. The DCA can
therefore not be used as a reliable method for quantitive predictions of those
observables.

There are two important factors that influence the results of the DCA, both
related to the choice of the cluster. The most obvious factor is the mean field
approximation, which reduces the momentum anisotropy of the self-energy as the
clusters become smaller. One can only avoid this error by considering clusters
with a sufficiently large size. In practice, the critical cluster-size is
obtained by comparing physical quantities on different cluster-sizes. More
complicated is the influence of the geometry of the cluster. There is a set of
different clusters, all of which have the same cluster size but different shape
and therefore different positions of the cluster momentum points. In
Fig.~\ref{fig:clusters}, we show two 16 site clusters for which this is the
case. The different positioning of the cluster momentum points in these two
clusters leads to a different geometric shape of the coarse-graining patches and
thus a different parametrization of the self-energy. This is illustrated in
Fig.~\ref{fig:sigma_band_structure_DCA}, where the momentum dependence of the
DCA self-energy at the lowest Matsubara frequency is shown for the 16A and 16B
site cluster introduced in Fig.~\ref{fig:clusters}. The relative error between
the self-energies on the different clusters is close to 100\% around the
Fermi-surface, making it unsuitable to derive any quantitative results from this
calculation.

One can argue that the influence of the mean-field approximation for clusters
with the same size is similar. Therefore, the difference in results can be
brought back to the shape of the coarse-graining patches. One example is the
difference in superconducting transition temperature $T_c$ between the 16A and
16B site cluster\cite{Maier2005PRL}. The role of the geometry has been studied
intensively by investigating the evolution of the magnetic and superconducting
transition temperatures over different cluster sizes\cite{Maier2005PRL,
Staar2012JPCS} or by comparing the site-occupancies of different clusters over a
wide range of doping \cite{Werner2009PRB, Gull2010PRB}.

The geometric shape dependence of the self-energy is built into the DCA by
construction, since the DCA self-energy is expanded on the coarse-grain patches
as\cite{Fuhrmann2007PRB}

\begin{equation}\label{DCA_Sigma}
\Sigma(\vec{k}, \varpi_m) = \sum_{i} \: \phi_{\vec{K}_i}(\vec{k})\,
\Sigma_{\vec{K}_i}(\varpi_m). \end{equation}

\noindent Here, the set of patches $\{\phi_{\vec{K}_i}(\vec{k})\}$ is formally
defined through the cluster-momenta $\{\vec{K}_i\}$,

\begin{equation}\label{patches} \phi_{\vec{K}_i}(\vec{k}) = \left\{
\begin{array}{rcl} 1 & \forall j : \vert\vec{k}-\vec{K}_i\vert \leq 
\vert\vec{k}-\vec{K}_j\vert \\ 0 &  \exists j : \vert\vec{k}-\vec{K}_i\vert > 
\vert\vec{k}-\vec{K}_j\vert \end{array}\right. \end{equation} and $\Sigma_{{\vec
K}_i}(\varpi_m)$ is the cluster self-energy for momentum ${\vec K}_i$.

In this paper, we present an extension to the DCA that allows the self-energy to
be expanded in an arbitrary large set of smooth basis-functions, and thereby
itself becoming a smooth function of momentum. The inclusion of a smooth
self-energy into the framework of the DCA requires a new fundamental look at the
algorithm. The resulting extended algorithm will be called \dcaplus, indicating
an incremental generalization to the well-known DCA algorithm. The
distinguishing feature of the \dcaplus algorithm that sets it apart from the DCA
algorithm is that cluster and lattice self-energies are in general different. 
In the DCA, the lattice self-energy $\Sigma(\vec{k})$ is a simple extension of
the cluster self-energy $\Sigma_{{\vec K}_i}$ via the step function form in
Eq.~(\ref{DCA_Sigma}). It therefore has jump discontinuities between the
patches. In the \dcaplus the lattice self-energy is a function with continuous
momentum dependence, which, when coarse grained is equal to the cluster
self-energy.

The focus of this paper is threefold. First, we will present the theoretical
background of the \dcaplus algorithm, without going into any practical details.
Next, we introduce a practical implementation for the \dcaplus algorithm and
discuss in detail the numerical aspects of the lattice mapping implementation.
Finally, we apply the \dcaplus algorithm to the single band Hubbard model in
order to investigate the pseudogap behavior, which has recently been
investigated in a systematic way with the DCA\cite{Werner2009PRB,Gull2010PRB}.
In the theory section, we first derive the coarse-graining equations for the
\dcaplus, which define how the lattice system is mapped onto an effective
cluster problem. This will introduce the key concepts of the \dcaplus approach
on a general level. Next, we discuss the structure of the \dcaplus algorithm in
more detail. Here, we will pay special attention to the lattice-mapping, i.e.
the inversion of the coarse-graining, where a lattice self-energy is estimated
from a given cluster self-energy. We will show that the lattice mapping is only
possible if the DCA assumption of a localized self-energy in real space is
uphold. In the implementation part, we will discuss the lattice-mapping in
detail on a practical level. In this paper, we propose to perform the lattice
mapping in two steps. First, we interpolate the self-energy obtained from the
cluster solver. Next, we deconvolute the interpolated cluster self-energy, where
the patch $\phi_{\vec{0}}(\vec{k})$ is used as the convolution kernel. In the
physics section, we will use the \dcaplus to investigate the pseudogap behavior
in the low-doping region of the two-dimensional Hubbard model. The self-energy
in this phase is known to be strongly momentum dependent and we will show that
the pseudogap transition temperature $T^*$ converges faster with regard to the
cluster size in the \dcaplus than in the DCA.

All calculations in this paper were performed for a single-band Hubbard model

\begin{equation}\label{eq:SBH} 
H=\sum_{ij} t_{ij}c^\dagger_{i\sigma}c^{\phantom\dagger}_{j\sigma} + U \sum_i
n_{i\uparrow}n_{i\downarrow}\,. \end{equation} 

Here $c^\dagger_{i\sigma}$ 
($c_{i\sigma}$) creates (destroys) an electron with spin $\sigma$ on lattice
site $i$ and $n_{i\sigma}=c^\dagger_{i\sigma}c^{\phantom\dagger}_{i\sigma}$ is
the corresponding number operator. The hopping matrix $t_{ij}$ includes nearest
($t=1$) and next-nearest ($t'$) neighbor hopping and $U$ is the on-site Coulomb
repulsion. The effective cluster problem of the DCA and \dcaplus is solved with
an continuous-time auxiliary-field quantum Monte Carlo algorithm
\cite{Gull2008,Gull2011}.

\section{Theory}

In this section, we present the generic structure of the \dcaplus algorithm,
without going into any implementation details. First, we introduce the key
features of the \dcaplus algorithm that distinguish it from the DCA, and show
that the latter is just a specialization of the former. Next, we present a
geometric interpretation of the \dcaplus algorithm in terms of the functional
representation space of the self-energy. This interpretation provides guidance
for how cluster-dependent features are incorporated into the lattice
self-energy, and offers insights for the derivation of a practical
implementation of the \dcaplus algorithm that will be discussed in the following
section. In order to keep the notation simple, we will omit the frequency
parameter $\varpi$ in all equations. Furthermore, all single-particle functions
defined on the impurity-cluster are represented by a subscript on the
cluster-momenta (e.g the cluster self-energy $\Sigma_{\vec{K}}$), while the
continuous lattice single-particle functions will have the usual dependence on
the momentum vector $\vec{k}$ (e.g. the lattice self-energy
$\Sigma(\vec{k})\:$). An overline over the quantity signifies that the latter
has been coarsegrained.

\subsection{DCA and \dcaplus formalisms:}

A system of interacting electrons on a lattice is generally described by a
Hamiltonian $H=H_0+H_{\rm int}$, where the kinetic energy $H_0$ is quadratic in
the fermion operators and the interaction $H_{\rm int}$ is quartic. It's free
energy $\Omega$ may be written in terms of the exact single-particle Green's
function $G$ as

\begin{equation}\label{eq:Omega} \Omega[G] = {\rm Tr}\ln (-{ G})+\Phi[G] -{\rm
Tr} [(G_0^{-1}-G^{-1})G] \,. \end{equation}

\noindent Here we have used a matrix notation for the Green's function $G$ of
the interacting system described by $H$ and the Green's function $G_0$ of the
non-interacting system described by $H_0$. $\Phi[G]$ is the Luttinger-Ward
functional\cite{Luttinger1960} given by the sum of all vacuum to vacuum
"skeleton" diagrams drawn with ${ G}$. The self-energy $\Sigma$ is obtained from
the functional derivative of $\Phi[G] $ with respect to $G$
\cite{Baym1961,Baym1962}

\begin{equation}\label{eq:SigmaFromPhi} \Sigma=\frac{\delta \Phi[G]}{\delta
G}\,, \end{equation}

\noindent and is related to the Green's function via the Dyson equation

\begin{equation}\label{eq:Dyson} G_0^{-1}-G^{-1}=\Sigma\,. \end{equation}

These two relations imply that the free energy is stationary with respect to
$G$, i.e. $\delta\Omega[G]/\delta G = 0$. In principle, the exact Green's
function $G$ and self-energy $\Sigma$ can be determined from the self-consistent
solution of Eqs.~(\ref{eq:SigmaFromPhi}) and (\ref{eq:Dyson}). However, since
the functional $\Phi[G]$ is usually unknown, an approximation is required that
replaces the exact $\Phi[G]$ by a known or a computable functional. Conserving
approximations replace the exact $\Phi[G]$ by an approximate functional, which
sums up certain subclasses of diagrams that are thought to capture the dominant
physics. In general, this results in a weak coupling approximation. A different
approach is taken in the DCA: rather than approximating the Luttinger Ward
$\Phi$, the functional representation space of the Green's function is reduced
by replacing the exact Green's function $G(\vec{k})$ by a
\textit{coarse-grained} Green's function $\bar{G}_{\vec{K}}$ in momentum space
defined as

\begin{equation}\label{eq:cgG} \bar{G}_{\vec{K}} = \int d\vec{k} \:
\phi_{\vec{K}}(\vec{k}) \: G(\vec{k})\,. \end{equation}

\noindent where the coarse-graining functions $\phi_{\vec{K}}(\vec{k})$ have
been defined in Eq.~(\ref{patches}).We note that approximating $G$ in this way
corresponds to an approximation of
the Laue function, $\Delta_{\vec{k}_1+\vec{k}_3,\vec{k}_2+\vec{k}_4}$, which
expresses momentum conservation at each vertex in the diagrams defining
$\Phi$\cite{Hettler2000PRB, Jarrell2001PRB}. For the single site DMFT
approximation ($N_c=1$), $\phi(\vec{k})$  is constant over the entire Brillouin
zone, and consequently the Laue function is replaced by $\Delta_{\rm DMFT}=1$,
i.e. momentum conservation is disregarded. For a finite size DCA cluster
($N_c>1$), the Laue function restores momentum conservation for the cluster
momenta $\vec{K}$ and reads in terms of the $\phi_{\vec{K}}(\vec{k})$

\begin{align}\label{DCALaue}
\Delta_{\rm{DCA}}(\vec{k}_1,\vec{k}_2,\vec{k}_3,\vec{k}_4) &=
\delta_{\vec{K}_1+\vec{K}_3,\vec{K}_2+\vec{K}_4}  \\ \times \,&
\phi_{\vec{K}_1}(\vec{k}_1)\, \phi_{\vec{K}_2}(\vec{k}_2) \,
\phi_{\vec{K}_3}(\vec{k}_3) \, \phi_{\vec{K}_4}(\vec{k}_4)\,. \nonumber
\end{align}

By replacing the exact Laue function with its DCA approximation in the Luttinger
Ward functional, the momentum integrals over the Green's functions in the
diagrams defining the $\Phi$-functional are reduced to sums over the finite set
of coarse-grained Green's functions defined in Eq.~(\ref{eq:cgG}). This way,
$\Phi[\bar{G}]$ becomes identical to the Luttinger-Ward functional of a finite
size cluster and the computation of the corresponding self-energy

\begin{align}\label{eq:DCAPhi}
\Sigma_{\vec{K}}^{\rm{DCA}}=\delta\Phi[\bar{G}_{\vec{K}}] /\delta
\bar{G}_{\vec{K}} \end{align}

\noindent becomes feasible. As such, within the DCA approximation the free
energy functional $\Omega[G]$ becomes

\begin{align} 
\Omega_{\rm DCA}[{ G}] = {\rm Tr}\ln (-{G})+\Phi[\bar{G}] -{\rm Tr} [(G_0^{-1}-G^{-1})G] \,. 
\nonumber 
\end{align}

From stationarity of the free energy, $\delta\Omega[G]/\delta G=0$, one obtains
the Dyson equation within the DCA

\begin{equation}\label{eq:DCADyson} G_0^{-1}(\vec{k}) - G^{-1}(\vec{k})=
\sum_{\vec{K}}\, \phi_{\vec{K}}(\vec{k}) \, \Sigma_{\vec{K}}^{\rm{DCA}}\,.
\end{equation}

\noindent Here, the right hand side follows from $\delta\bar{G}_{\vec{K}} /
\delta G= \phi_{\vec{K}}(\vec{k})$ and $\delta
\Phi[\bar{G}_{\vec{K}}]/\delta\bar{G}_{\vec{K}}=\Sigma_{\vec{K}}^{\rm{DCA}}$.
Eqs.~(\ref{eq:cgG}), (\ref{eq:DCAPhi}) and (\ref{eq:DCADyson}) form a closed set
of equations which is solved iteratively until self-consistency is reached. This
is the DCA algorithm. Following Eq.~(\ref{eq:DCADyson}), the self-energy
$\Sigma(\vec{k})$ of the lattice Green's function $G(\vec{k})$, which is used to
compute the coarse-grained Green's function in Eq.~(\ref{eq:cgG}), is
approximated by a piecewise constant continuation of the cluster self-energy
$\Sigma_{\vec{K}}^{\rm{DCA}}$, which changes between different momentum patches
but is constant within a given patch,

\begin{equation} \label{eq:DCASigma} \Sigma(\vec{k}) =
\sum_{\vec{K}}\,\Sigma^{\rm{DCA}}_{\vec{K}} \,\phi_{\vec{K}}(\vec{k}).
\end{equation}

With the \dcaplus algorithm we introduce in this paper, the DCA framework is
extended to allow for a more general relationship between the lattice
self-energy $\Sigma(\vec{k})$ and cluster self-energy $\Sigma_{\vec{K}}$ than
that in  Eq.~(\ref{eq:DCASigma}). In the \dcaplus, in analogy with
Eq.~(\ref{eq:cgG}), we only demand the cluster self-energy to be equal to the
coarse-grained lattice self-energy,

\begin{equation}\label{eq:cgSigma} \bar{\Sigma}_{\vec{K}} = \int d\vec{k}\,
\phi_{\vec{K}}(\vec{k})\, \Sigma(\vec{k})\,. \end{equation}

In the DCA algorithm, this requirement is trivially satisfied since according to
Eq.~(\ref{eq:DCASigma}), $\Sigma(\vec{k})$ is set to the cluster self-energy
$\Sigma(\vec{K})$ for momenta $\vec{k}$ in patch $P_i$. However, it is important
to realize that Eq.~(\ref{eq:cgSigma}) allows for a more general approximation
of the lattice $\Sigma(\vec{k})$, which, for example, can retain its smooth
momentum dependence instead of the DCA step function character. To proceed, it
is convenient for our purposes to express the free energy as a functional of the
self-energy. By following the work of
Potthoff\cite{Potthoff2003,Potthoff2003EPJ}, we eliminate the Green's function
$G$  in favor of the self-energy $\Sigma$ to write the free energy as a
functional of the self-energy $\Sigma$,

\begin{equation}\label{eq:OmegaOfSigma} \Omega[\Sigma] =  -{\rm Tr}\ln
[-(G_0^{-1}-\Sigma)] + (\mathcal{L} \Phi)[\Sigma]\,. \end{equation}

\noindent Here, the functional $(\mathcal{L} \Phi)[\Sigma]$ is obtained from
$\Phi[G]$ through a Legendre-transformation

\begin{equation}\label{eq:Legendre} (\mathcal{L} \Phi)[\Sigma] = \Phi - {\rm
Tr}[\Sigma\,G]\,. \end{equation}

Replacing $\Sigma(\vec{k})$ in $(\mathcal{L} \Phi)[\Sigma]$ with the
coarse-grained self-energy in Eq.~(\ref{eq:cgSigma}), i.e. $\Sigma(\vec{k})
\approx \sum_{\vec{K}} \phi_{\vec{K}}(\vec{k})\bar{\Sigma}_{\vec{K}}$, then
yields

\begin{equation} \label{eq:phiDCAp} (\mathcal{L} \Phi)[\Sigma] = \Phi -
\sum_{\vec{K}} \: \bar{\Sigma}_{\vec{K}}\, \bar{G}_{\vec{K}}\,, \end{equation}

\noindent where $\bar{G}_{\vec{K}}$ is the coarse-grained Green's function
defined in Eq.~(\ref{eq:cgG}). If this functional is used in the free energy in
Eq.~(\ref{eq:OmegaOfSigma}), one obtains at stationarity,
$\delta\Omega[\Sigma]/\delta\Sigma=0$,

\begin{equation} [G_0^{-1}(\vec{k})-\Sigma(\vec{k})]^{-1} = \sum_{\vec{K}}
\,\phi_{\vec{K}}(\vec{k}) \, \bar{G}_{\vec{K}}\,. \end{equation}

\noindent Here, the right hand side follows from
$\delta\bar{\Sigma}_{\vec{K}}/\delta \Sigma = \phi_{\vec{K}}(\vec{k})$ and
$(\mathcal{L}
\Phi)[\bar{\Sigma}_{\vec{K}}]/\delta\bar{\Sigma}_{\vec{K}}=-\bar{G}_{\vec{K}}$.
Using the identity $\int d\vec{k}
\phi_{\vec{K}}(\vec{k})\phi_{\vec{K}'}(\vec{k}) = \delta_{\vec{K},\,\vec{K}'}$
and multiplying both sides with $\int d\vec{k} \phi_{\vec{K}}(\vec{k})$ results
in the \dcaplus coarse-graining equation

\begin{equation}\label{eq:cgGDCAp} \bar{G}_{\vec{K}} = \int d\vec{k} \,
\phi_{\vec{K}}(\vec{k}) \, [G_0^{-1}(\vec{k})-\Sigma(\vec{k})]^{-1}\,.
\end{equation}

We note that in contrast to the DCA algorithm, the lattice self-energy
$\Sigma(\vec{k})$ enters in the coarse-graining step. It is related to the
cluster self-energy $\Sigma_{\vec{K}}$ through Eq.~(\ref{eq:cgSigma}), i.e. its
coarse-grained result must be equal to $\Sigma(\vec{K})$. The special choice
$\Sigma(\vec{k}) = \sum_{\vec{K}}\phi_{\vec{K}}(\vec{k})\Sigma_{\vec{K}}$
satisfies this requirement and recovers the DCA algorithm. But in general,
$\Sigma(\vec{k})$ needs to only satisfy Eq.~(\ref{eq:cgSigma}), i.e. one has
more freedom in determining a lattice self-energy $\Sigma(\vec{k})$ from the
cluster $\Sigma(\vec{K})$. In the \dcaplus algorithm, we take advantage of this
freedom to derive a $\Sigma(\vec{k})$ that retains a smooth $\vec{k}$-dependence
and thus is more physical than the piecewise constant $\Sigma(\vec{k})$ of the
DCA. As in the DCA, the cluster self-energy $\Sigma_{\vec{K}}$ may be determined
from the solution of an effective cluster problem described by $({\cal
L}\Phi)[\Sigma]$ as a functional of the coarse-grained propagator
$\Sigma[\vec{K}]=\Sigma[\bar{G}(\vec{K})]$. This, together with
Eqs.~(\ref{eq:cgSigma}) and (\ref{eq:cgGDCAp}) form the basis of the \dcaplus
algorithm.

A detailed description of the algorithm will be given in the implementation
section. Evidently, determining the lattice self-energy $\Sigma(\vec{k})$ from
the cluster self-energy $\Sigma_{\vec{K}}$ through inversion or deconvolution of
Eq.~(\ref{eq:cgSigma}) presents a difficult task.

\subsection{Structure of a \dcaplus cluster-calculation:}

Since the lattice self-energy $\Sigma(\vec{k})$ no longer is restricted to
Eq.~(\ref{DCA_Sigma}), it can be expanded into an arbitrary set of smooth basis
functions $\{\mathcal{B}_{i}(\vec{k})\}$, such as cubic splines or crystal
harmonics, i.e. \begin{equation}\label{eq:Sigma_expansion} \Sigma({\vec k}) =
\sum_i {\cal B}_i({\vec k})\sigma_i\,. \end{equation} Here, $\sigma_{j}$ are the
expansion coefficients of the lattice self-energy corresponding to the
basis-function $\mathcal{B}_{j}(\vec{k})$. Contrary to the DCA, the
coarse-graining patches $\phi_{\vec{K}}(\vec{k})$ in the \dcaplus are not linked
in any shape or form to the basis functions in which we expand the lattice
self-energy. As was mentioned in the previous section, the \dcaplus maps the
full lattice problem into a cluster impurity problem embedded into a mean field
by coarse-graining both the lattice self-energy and lattice Green's function.
The cluster-mapping in the \dcaplus is thus very similar to the cluster-mapping
in the DCA, with the exception that we use a continuous lattice self-energy in
the coarse-graining of the Green's function

\begin{align}\label{cluster_mapping}
\bar{\Sigma}_{\vec{K}} &=\frac{N_c}{V_{BZ}}
\int_{BZ} d\vec{k} \: \phi_{\vec{K}}(\vec{k})\: \Sigma(\vec{k}),  \\
\bar{G}_{\vec{K}} &= \frac{N_c}{V_{BZ}} \int_{BZ} d\vec{k} \:
\phi_{\vec{K}}(\vec{k})\:\: \Big[ [G^0(\vec{k})]^{-1}-\Sigma(\vec{k})
\Big]^{-1}. \nonumber \end{align}

Eq.~(\ref{cluster_mapping}) can now be simplified by using the explicit
expansion of the lattice self-energy in Eq.~(\ref{eq:Sigma_expansion})
\begin{align}\label{cluster_mapping_2} \bar{\Sigma}_{\vec{K}_i} &=  \sum_{j}
\underbrace{\: \Bigg( \int \: d\vec{k} \: \phi_{\vec{K}_i}(\vec{k}) \:
\mathcal{B}_{j}(\vec{k}) \Bigg) }_{=\: P_{i,j}}\,  \sigma_{j}. \end{align} Here,
$P_{i,j}$ is a projection operator, defined by coarse-graining the basis
function $\mathcal B_j$ over  patch $i$. Note that in the DCA, this projection
operator becomes the identity-operation $\delta_{i,j}$. Hence, the coarse
graining of the lattice self-energy in the DCA is an implicit operation
($\sigma_{i} \equiv \bar{\Sigma}_{\vec{K}_i}$), while in the \dcaplus it becomes
explicit.

With the introduction of the cluster-mapping in the \dcaplus in
Eq.~(\ref{cluster_mapping_2}), the lattice mapping is conceptually well defined
as long as the inverse of the projection-operator $P$ exists. Assuming that
$P^{-1}$ exists, we can retrieve the expansion coefficients of the lattice
self-energy from the self-energy of the cluster-solver $\Sigma_{\vec{K}}$ in a
straightforward manner

\begin{align}\label{lattice_mapping} \sigma_{j} &=  \sum_{j}  \: (P^{-1})_{i,j}
\: \Sigma_{\vec{K}_j}. \end{align}

\begin{figure}[t] \begin{center}
\includegraphics[width=0.5\textwidth]{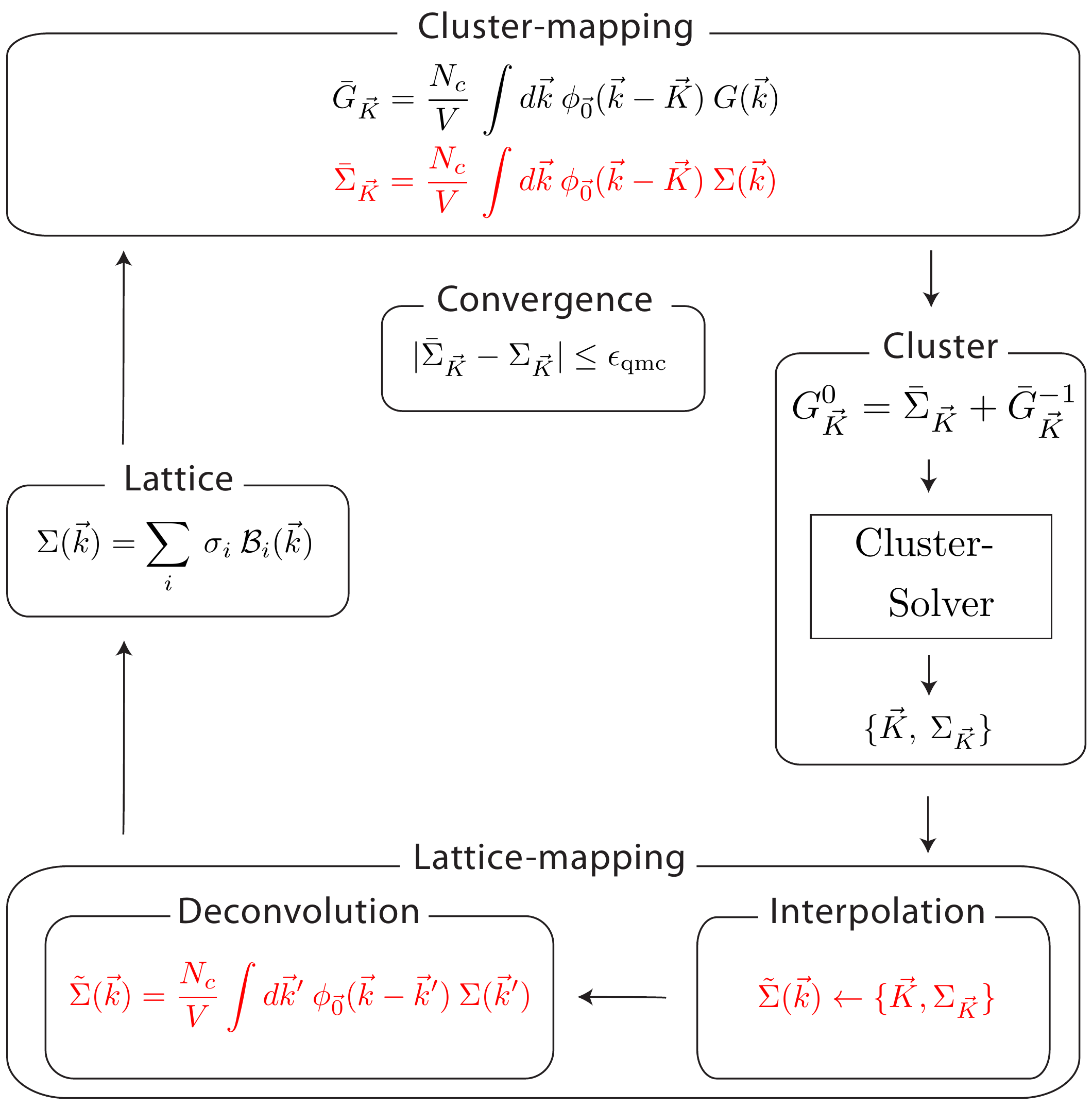} \end{center}
\caption{\label{fig:DCAPLUSdiagram} The generic structure of a self-consistent
DCA$^{+}$ algorithm, in which the cluster- and lattice-mapping play a central
role in order to connect the continuous lattice self-energy $\Sigma(\vec{k})$
with the cluster self-energy $\Sigma_{\vec{K}}$. Convergence is reached when the
cluster-solver produces a cluster self-energy $\Sigma_{\vec{K}}$ equal to the
coarse-grained self-energy $\bar{ \Sigma}_{\vec{K}} \equiv \bar{
\Sigma}(\vec{K})$. } \end{figure}

This closes the \dcaplus iteration and allows us to carry out a self-consistent
calculation.

In Fig.~\ref{fig:DCAPLUSdiagram}, we have summarized the generic structure of
the DCA$^{+}$ algorithm, without specifying yet any implementation details of
the lattice-mapping. In the "cluster-mapping" step, the lattice Green's function
and self-energy are coarse-grained onto the patches defined by $\Phi_{\vec
K}({\vec k})$ to give ${\bar G}_{\vec{K}}$ and ${\bar \Sigma}_{\vec{K}}$,
respectively. A cluster solver algorithm such as QMC is then used to calculate,
from the corresponding bare Green's function $G_{0,{\vec{K}}}$, the interacting
Green's function and self-energy $\Sigma_{\vec K}$ on the cluster. In the
"lattice-mapping step", which is missing in the standard DCA algorithm, a new
estimate for the lattice self-energy $\Sigma({\vec k})$ is then computed through
inversion of the projection operator $P_{i,j}$. The lattice self-energy then
enters the next cluster-mapping step via the lattice Green's function $G({\vec
k})$. In the implementation section of this paper, we will describe in detail
how the lattice-mapping can be done in a numerically stable way.

Due to the distinction between the lattice and cluster self-energy in the
\dcaplus algorithm, we can not use the convergence criteria of the DCA. In the
latter, convergence is reached if the self-energy (lattice or cluster) of the
previous iteration is equal to the current one. If one monitors only convergence
on the lattice self-energy in the \dcaplus algorithm, one might stop the
iterations although the cluster solver still produces a cluster self-energy
$\Sigma_{\vec{K}}$ that differs from the coarse-grained lattice self-energy
$\bar{\Sigma}(\vec{K})$. This would indicate that the \dcaplus does not converge
to a stationary point of the free energy functional $\Omega$. To avoid such a
problem, we demand that convergence is reached only when the coarse-grained
lattice self-energy $\bar{\Sigma}(\vec{K})$ and the cluster self-energy
$\Sigma_{\vec{K}}$ agree to within the Monte Carlo sampling error.

It is important to note that the proposed algorithm is fundamentally different
from a simple interpolation of the cluster self-energy $\Sigma_{\vec K}$ between
the cluster momenta ${\vec K}$. A smooth interpolation will almost certainly
fail to satisfy Eq.~(\ref{eq:cgSigma}), i.e. the main requirement of the \dcaplus
that the coarse-grained lattice $\Sigma({\vec k})$ is equal to the cluster
$\Sigma_{\vec K}$. Such a procedure was proven in Ref.~\cite{Hettler2000PRB} to
lead to causality violations when the cluster self-energy is added back to the
inverse coarse-grained propagator in the "cluster exclusion" step to avoid
overcounting of self-energy diagrams. In the \dcaplus, the lattice self-energy is
different from an interpolated cluster self-energy and the self-energy that
enters the cluster exclusion step is given by the coarse-grained lattice
self-energy. Because of this, the proof given in Ref.~\cite{Hettler2000PRB} does
not apply and the \dcaplus algorithm is not automatically plagued by causality
problems. Although we do not have a rigorous proof that the \dcaplus algorithm
remains causal, we have never encountered any causality violations in the
application of this method to the single-band Hubbard model.

The projection operator $P_{i,j}$ plays a central role in the implementation of the \dcaplus 
algorithm. In order to obtain a self-consistent algorithm, it is conceptually clear that the
projection operator has to be invertible. In practice, however, this might not be straightforward to
achieve. An intuitive understanding of this operator is developed in Appendix 1, where we discuss how the projection operator influences the choice of the cluster, and we show that
its inverse only exists if the DCA locality assumption for the lattice self-energy is satisfied.

\subsection{Role of the cluster in the \dcaplus}

In the DCA algorithm, the real space cluster takes a central role. It completely
defines the basis-functions in which the self-energy is expanded. Furthermore,
the real space cluster dictates how the lattice is mapped on the cluster through
the coarse-graining procedure. Consequently, solutions obtained with the DCA
algorithm usually dependent on the particular choice (shape) of the cluster. In
practice, this leads to a very good qualitative description of the physics, but
prohibits quantitative analysis, as calculated physical quantities strongly
depend on cluster shape. In the \dcaplus, we start from an expansion of the
self-energy into an arbitrary set of basis-functions. In this way, the influence
of the real space cluster is reduced, since it does not dictate the
basis-functions on which the self-energy is expanded. The real space cluster
only specifies how the cluster is mapped on the lattice through the shape of the
coarse-graining patches. Consequently, the focus in the \dcaplus shifts from the
real space cluster to the projection operator $P_{i,j}$. This operator embodies
the quantum cluster approximation of the \dcaplus, since it connects the cluster
self-energy with the lattice self-energy in a purely geometric way. The
projection operator is only defined by the set of basis-functions of the lattice
self-energy and the real space cluster and not subjected in any way to physical
parameters (such as temperature, band-structure, interaction terms, ...). This
purely 'geometric' property of the projection operator allows us to find a
priori the necessary conditions to which the cluster self-energy has to be
subjected, in order to allow for a self-consistent, cluster-independent \dcaplus
calculation. These necessary conditions that follow from the discussion in the
previous subsections and Appendix 1 are:

\begin{itemize} 
\item In order to perform a self-consistent \dcaplus
calculation, the cluster self-energy has to converge in the image-space
$\mathcal{I}_{\epsilon}$ of the projector. 
\item In order to perform a
cluster-independent \dcaplus calculation on the cluster $A$ and $B$, the cluster
self-energy needs to converge on the intersection of the image-spaces of both
projectors ($\mathcal{I}^A_{\epsilon} \bigcap \mathcal{I}^B_{\epsilon}$) .
\end{itemize}

\section{implementation}

In the last section, we have introduced a projection operator $P_{i,j}$ and
shown its involvement in the cluster and lattice-mapping.  Via a geometric
consideration, we have shown conceptually that its inverse exists as long as the
expansion coefficients $\langle \bar{\Sigma}_{\vec{k}}, e_{\lambda}(\vec{k})
\rangle$ of the cluster self-energy vanish rapidly in the image-space
$\mathcal{I}_{\epsilon}$ of the projection operator $P_{i,j}$. At closer
inspection, the lattice mapping is thus a two stage process. First, we need to
determine the expansion coefficients of the cluster self-energy. To this end, we
will propose a novel interpolation technique, which is motivated from the
analytical properties of the self-energy. The interpolated cluster self-energy
${\bar \Sigma}_{{\vec k}_j}$ is then used to compute the inner product $\langle
{\bar \Sigma}_{{\vec k}_j},e_\lambda({\vec k}_j)\rangle$ with the eigenfunctions
of the projection operator $P_{i,j}$, which gives the expansion coefficients of
the cluster self-energy. Secondly, we need to deconvolute the interpolated
cluster self-energy on the image space $\mathcal{I}_{\epsilon}$, where we need
to determine the optimal value for the parameter $\epsilon$. If the latter is
too large, the self-consistency can not be reached. If $\epsilon$ is too small,
the lattice-mapping will become numerically unstable due to the division of
small eigenvalues. To solve this problem, we adapt the Richardson-Lucy
deconvolution algorithm, which inverts Eq.~(\ref{cluster_mapping_2}) in a numerically
stable way.

\subsection{Interpolation}

\begin{figure}[t] \includegraphics[width=0.5\textwidth]{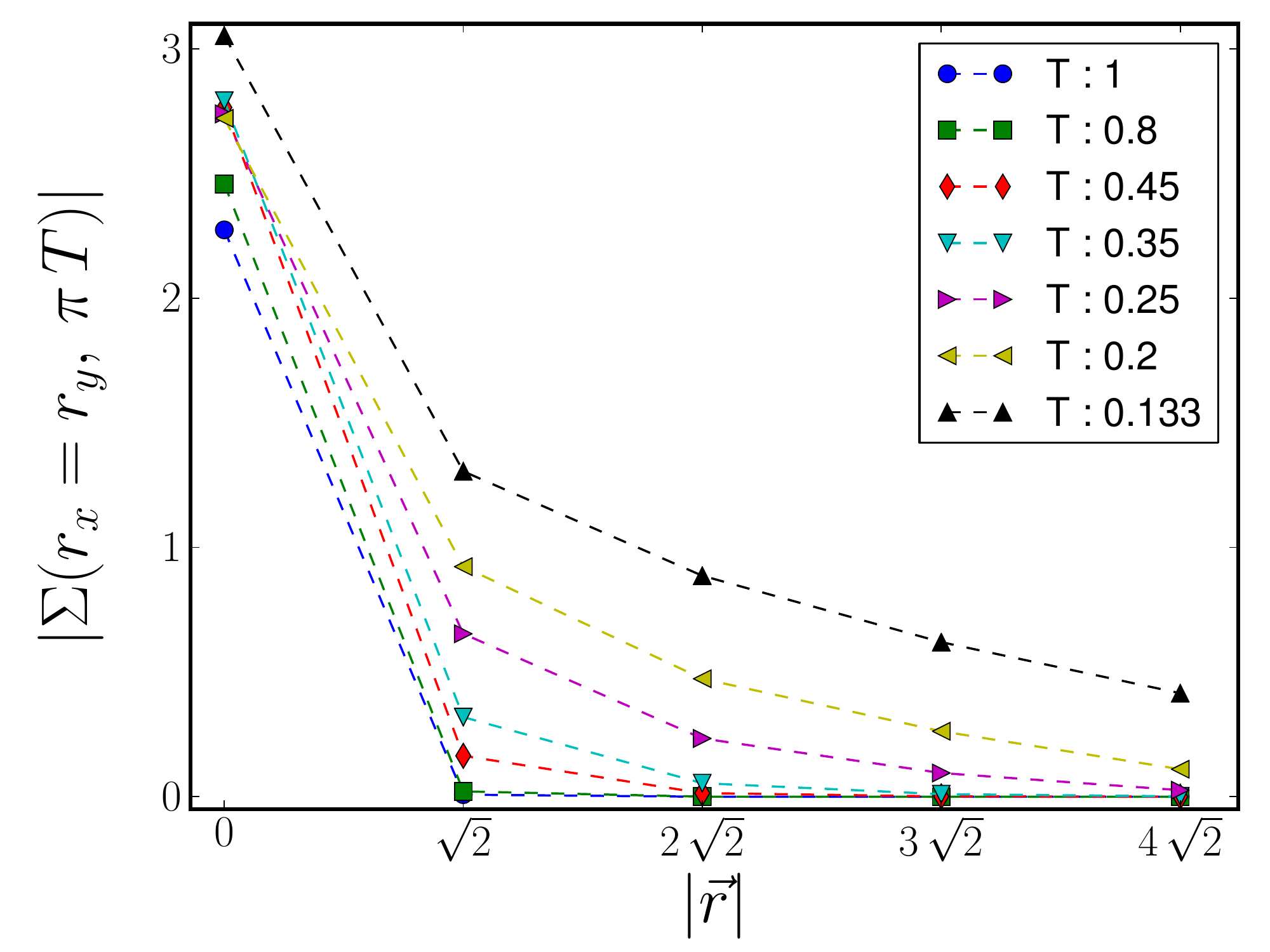}
\caption{\label{fig:Sigma_Nc100_decay} The decay of $\Sigma_R$ for a
$N_c=100$-site cluster with $U/t=7$ and $t'/t=0$ for various temperatures at
half filling. For high temperatures ($T\leq0.3$), the system is only weakly
correlated and $\Sigma_R$ will rapidly decay. For low temperatures, the
correlations exceed the cluster-radius $R_c=5$.   } \end{figure}

In the context of tight-binding models, one of the most successful algorithms to
interpolate its band structure is the
Wannier-interpolation-method\cite{Vanderbilt2011RMP}. It finds its justification
in the localized nature of Wannier orbitals, from which the tight-binding models
are derived. Since the self-energy is a correction to the band-structure due the
interaction between the electrons, the Wannier interpolation method seems a
suitable interpolation algorithm. Okamoto et al.~\cite{Okamoto2003PRB} have
examined this possibility implicitly, by expanding the lattice self-energy
$\Sigma(\vec{k})$ into the cubic-harmonic basis-functions
$\{\mathcal{C}_{\vec{K}}(\vec{k})\}$.

\begin{align}\label{Okamoto_eqns_1} \Sigma(\vec{k}) &=  \sum_{\vec{K}}
\mathcal{C}_{\vec{K}}(\vec{k}) \: \Sigma_{K} \\ \mathcal{C}_{\vec{K}}(\vec{k})
&= \frac{1}{N_c} \sum_{\vec{R}} e^{\imath \vec{R} (\vec{K}-\vec{k}) } \nonumber
\end{align}

This approach only works when the self-energy $\Sigma_{\vec{K}}$ is sufficiently
smooth, such that the real-space self-energy $\Sigma_{\vec{R}}$ converges on the
cluster in real space. Notice that the latter is implicitly computed in
Eq.~(\ref{Okamoto_eqns_1}), since

\begin{align} \Sigma(\vec{k}) &=  \sum_{\vec{K}} \mathcal{C}_{\vec{K}}(\vec{k})
\: \Sigma_{\vec{K}} =  \sum_{\vec{R}} e^{-\imath \vec{R}\, \vec{k} }  \:
\underbrace{\frac{1}{N_c}  \sum_{\vec{K}} e^{\imath \vec{R} \vec{K} } 
\Sigma_{\vec{K}}}_{=\Sigma_{\vec{R}} } .  \nonumber \end{align}

The sum over all lattice points can now be split into two terms. In the first
term, we run over all lattice-points within the cluster-radius. In the second
term, we sum over all the remaining points in the lattice.

\begin{align}\label{directapproach} \Sigma(k) &= \sum_{\vec{R}} e^{-\imath
\vec{R} \vec{K} }  \Sigma_{\vec{R}} \nonumber \\ &=  \sum_{\vert \vec{R} \vert <
R_c} e^{-\imath \vec{R} \, \vec{k} }  \Sigma_{\vec{R}} + \sum_{\vert \vec{R}
\vert \geq R_c} e^{-\imath \vec{R} \, \vec{k} }  \Sigma_{\vec{R}} \end{align}

If correlations have longer range, $\Sigma_{\vec{R}_i}$ will no longer converge
on the  cluster in real space.  This is clearly illustrated in
Fig.~\ref{fig:Sigma_Nc100_decay}, where we show the self-energy
$\Sigma_{\vec{R}} $ for a $N_c=100$-site cluster with $U/t=7$ for various
temperatures. At  high temperatures ($T\geq0.25$), the system is only weakly
correlated. The self-energy  $\Sigma_{\vec{R}}$ in this temperature range is
contained within the cluster-radius $R_c=5$. For lower temperatures, it is clear
that $\Sigma_{R}$ is extends beyond $R_c$. Applying the Wannier interpolation
scheme according to Eq.~(\ref{Okamoto_eqns_1}) to such correlated systems is
simply not allowed, since the expansion coefficients $\Sigma_{\vec{R}}$ outside
the cluster can not be assumed to be zero. A straightforward application of
Eq.~(\ref{Okamoto_eqns_1}) will lead to ringing and eventually to causality
violations. The latter was observed by Okamoto et al.~\cite{Okamoto2003PRB}, and
could only partially be resolved by introducing low-pass filtering schemes. The
applicability of this approach is very limited, due to a lack of a general
framework to determine these filters.

\subsubsection{Formalism of the interpolation:}

From the previous section, it has become clear that the interpolation techniques
such as Eq.~(\ref{Okamoto_eqns_1}) can only work if the function converges on
the finite (and often small) basis-set. The rate of convergence depends
critically on the choice of the basis-functions. Consider for example the free
Green's function $G^0$ of the single band Hubbard model in Eq.~(\ref{eq:SBH}),

\begin{align}\label{G0} G^0(\vec{k}, \varpi)  = [\imath\:\varpi +
\epsilon(\vec{k})]^{-1}. \end{align}

\noindent While this Green's function $G^0$ will converge poorly on the
cubic-harmonics of the lattice for small frequency $\varpi$, it is
straightforward to see that  $[G^0]^{-1}$ will be completely converged on a
4-site cluster. This simple example shows how one can extend the
interpolation-idea introduced by Okamoto et al\cite{Okamoto2003PRB}. Given an
injective transformation $\mathcal{T}$, we can write

\begin{align}\label{indirect approach} \mathcal{F}(\vec{k}) &=
\mathcal{T}^{-1}\Big[ \mathcal{T}\big[ \mathcal{F}\big](\vec{k}) \Big]\nonumber
\\ &= \mathcal{T}^{-1}\Big[ \sum_{\vec{K}} \mathcal{C}_{\vec{K}}(\vec{k}) \: 
\mathcal{T}\big[ \mathcal{F}_{\vec{K}} \big] \Big] \end{align}

\begin{figure*}[!] \begin{center}
\includegraphics[width=\textwidth]{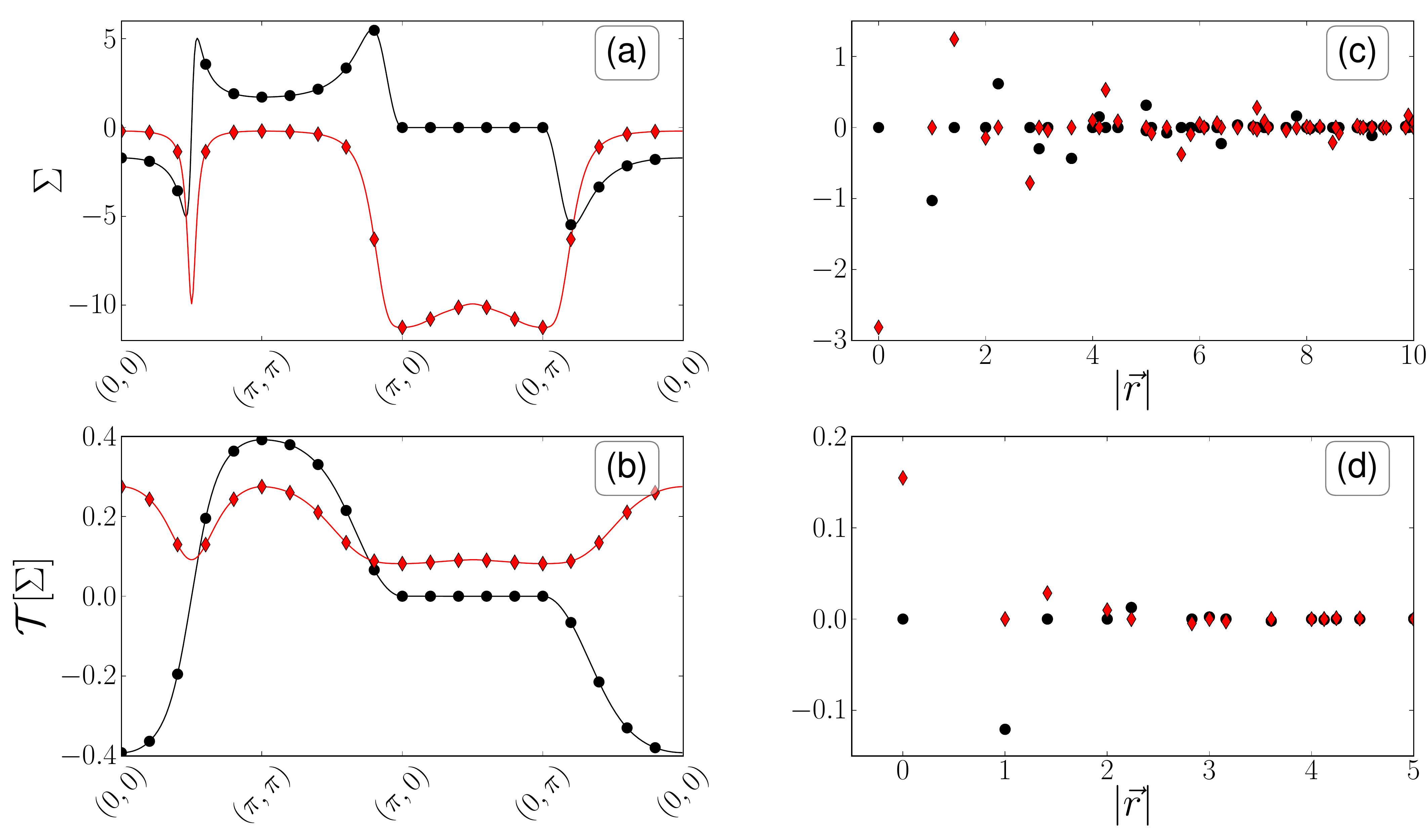}
\end{center} \caption{\label{fig:overview_Sigma_Nc100} Interpolation-procedure
for the self-energy $\Sigma_K$ at the lowest matsubara frequency for a
$100$-site cluster at a temperature $T=0.2$, $U/t=7$ and $t'=0$ at half-filling.
(A) The interpolated function $\Sigma(\vec{k})$ is a smooth function through the
results $\Sigma_K$ obtained from the QMC cluster solution, where the circles and
diamonds represent respectively the real and imaginary part. (B) The transformed
function $\mathcal{T}[\Sigma]$ smoothes the self-energy function, making it
suitable for a cubic harmonics expansion. (C) The Fourier transform of the
interpolated function $\Sigma(k)$. Notice that the tails expand much further
than the cluster-radius $R_c=5$. (D) The Fourier transform of the function
$\mathcal{T}[\Sigma_K]$. The convergence is reached at $R_c=3$.} \end{figure*}

The method of operation to interpolate a function becomes now clear. Find an
injective (and preferably analytical) transformation $\mathcal{T}$, such that
the transformed  function-values converge on the chosen basis-functions. Use
this expansion to compute the transformed function-values on arbitrary k-points.
Finally, apply the inverse transformation $\mathcal{T}^{-1}$ on the transformed
function-values in order to obtain the desired interpolated function-values on
arbitrary k-points.

This approach has many advantages. First, it provides a measure that indicates
when the interpolation-procedure works or fails. If $\mathcal{T}[
\Sigma_{\vec{K}}]$ does not converge on the basis-set, one is not allowed to
perform an interpolation. Second, this interpolation-procedure does not
introduce \textit{extra information} \--- filtering schemes and other numerical
tricks to ensure causality, on the other hand, introduce extra, undesirable
structure into the interpolated functions. By using filtering schemes or other
numerical tricks to assure causality, we introduce extra structure in the
function that is to be interpolated, which is undesirable. Third, if the
transformation $\mathcal{T}$ is analytical, we will not break the analyticity of
the interpolated function. For Green's functions and their derived functions
such as the self-energy, analyticity is an important property. In arbitrary
interpolation schemes such as splines\cite{Boor1978Splines} or radial-basis
expansions\cite{Buhmann2003RBF}, this analyticity is often broken. The obtained
interpolating function is therefore questionable from a physics-point of view.
The challenge of this approach is naturally the search for a correct
transformation $\mathcal{T}$. Notice that $\mathcal{T}$ can be different for
different functions, since the only requirements are injectivity and convergence
on the chosen basis-set. In the next subsections, we will propose such a
transformations for the self-energy $\Sigma$. The proposed transformation will
be motivated by physical and analytical properties of the self-energy.

\subsubsection{Interpolation on large clusters:}

Since the imaginary part of the self-energy is strictly negative in the
upper-half of the complex-plane~\cite{AGD} \begin{align}\label{Sigma_im_prop}
\mbox{Im}[\Sigma(\vec{k},\imath \: \varpi > 0)] < 0. \end{align} we can
introduce an injective transformation $\mathcal{T}$ that preserves the
analyticity of the self-energy\footnote{Since $
\mbox{Im}[\Sigma(\vec{k},\varpi)] < 0$, we will not introduce any new poles in
the upper-half plane by inverting the function.},

\begin{align} \mathcal{T}(\Sigma) = \big[ \Sigma-\alpha\,\imath\big]^{-1}, \:
\rm{with} \:\:  \alpha > 0. \end{align}

\noindent Due to the property shown in (\ref{Sigma_im_prop}), the transformation
$\mathcal{T}$ will map the self-energy $\Sigma$ into a bounded function,
irrespective of how spiky the self-energy $\Sigma$ is. Notice also that we first
shift the imaginary part of the self-energy down by $\alpha\, \imath$, in order
to avoid introducing poles due to the Monte Carlo statistical noise. 
Consequently, the function $\mathcal{T}(\Sigma)$ will now be localized in real
space, and we can safely perform an expansion of the function
$\mathcal{T}(\Sigma)$ over cubic harmonics. We have illustrated this process in
Fig.~\ref{fig:overview_Sigma_Nc100}, by applying our interpolation procedure to
a $100$-site cluster at a temperature $T=0.2$ at half filling.  In (A), we show
respectively the computed values of the cluster self-energy $\Sigma_{\vec{K}}$
and its interpolation $\Sigma_{\vec{k}}$ along a high-symmetry line in the
Brillouin-zone. Notice that the imaginary part of the interpolation function
remains at all times negative! In (B), the transformed function
$\mathcal{T}[\Sigma_{\vec{K}}]$ is shown, together with its interpolating
function. Clearly, the transformation $\mathcal{T}$ has reduced the sharp
features in the self-energy, and the function has become smoother. 
In (C) and (D), we show the Fourier transform from respectively the interpolated
self-energy $\Sigma(\vec{k})$ and the transformed values
$\mathcal{T}[\Sigma_{\vec{K}}]$. The large difference in the convergence radii
is clear, and shows the effectiveness of our indirect approach compared to a
direct one. This result is not a coincidence. In the Appendix 2, we have proven in
a rigorous way the point-wise convergence.

\subsubsection{Interpolation on small clusters:}

For certain parameter sets, the fermionic sign problem prevents the
investigation of large enough clusters, for which $\mathcal{T}[\Sigma]$ will
converge. In this case, we recommend to interpolate the $\mathcal{T}[\Sigma]$
using cubic splines, instead of interpolating the latter with the earlier proposed Wannier-interpolation.
Since $\mathcal{T}[\Sigma]$ is a much smoother function, cubic splines can still
perform reasonably well, even in the case of small clusters. The self-energy on
the other hand will not be smooth, and a straightforward spline interpolation
will lead to overshoots or ringing, which in turn turn might lead to an
acausal self-energy. This particular phenomenon has been studied extensively by
Okamoto et al\cite{Okamoto2003PRB}. The ringing might be cured by the use of
tension splines\cite{Cline1974}, in which case a tension parameter is
introduced. It is however important to keep in mind that the splines might add
extra information into the system, and thus bias the physics. This problem does
not occur with Wannier interpolation, as long as the Fourier coefficients of
$\mathcal{T}[\Sigma_{\vec{K}}]$ converge on the real space cluster.

\subsubsection{lattice-symmetry:}

Most of the clusters used in the DCA do generally not obey the same symmetry 
operations as the infinite lattice. As a consequence, the lattice-self-energy in 
the DCA breaks the symmetry of the lattice, due to its strict parametrization 
with the coarsegrain patches. The only way to resolve this issue in the DCA, is to restrict 
to the few clusters that obey the cluster-symmetry. In order to remove this 
undesirable feature in the \dcaplus, we symmetrize the self-energy after the 
interpolation. The interpolated cluster-self-energy obeys thus by construction 
the symmetry operations of the lattice.

\subsection{Cluster Deconvolution}

The goal of this section is to present a practical implementation of the
lattice-mapping. As mentioned in the theoretical section of this paper, the
lattice mapping is in essence the inversion of the cluster mapping defined in
Eq.~(\ref{cluster_mapping_2}). In a common \dcaplus calculation, we will have much
more basis functions than Monte Carlo cluster-points. As a consequence, we need
to determine more lattice expansion coefficients than cluster-points that are
given by the cluster-solver. The inversion problem is thus seemingly
underdetermined. Therefore, we do not attempt to invert
Eq.~(\ref{cluster_mapping_2}) directly, but first generalize the coarsegraining equation
of the self-energy. This is accomplished by rewriting each coarsegraining patch as a
translation of the patch around the origin, i.e $\phi_{\vec{K}}(\vec{k}) = 
\phi_{\vec{0}}(\vec{k}-\vec{K})$. Next, we generalize the cluster-momentum vector $\vec{K}$
to an arbitrary momentum vector. Using the interpolated cluster self-energy 
$\bar{\Sigma}_{\vec{K}}$ as a substitute for the cluster self-energy $\Sigma_{\vec{K}}$ 
in Eq.~(\ref{cluster_mapping}), we obtain

\begin{align}\label{cluster_convolution} 
\bar{\Sigma}(\vec{k}) &=\frac{N_c}{V} \int d\vec{k'} \: \phi_{\vec{0}}(\vec{k}-\vec{k'})\: 
\Sigma(\vec{k'})
\end{align}

Any solution of Eq.~(\ref{cluster_convolution}) is thus also a solution of
Eq.~(\ref{cluster_mapping_2}). We should stress that with the exception of the
continuity of the self-energy, this generalization does not introduce any new
information as long as the Wannier-interpolation converges! With
Eq.~(\ref{cluster_convolution}), we have now rephrased the lattice-mapping into
a deconvolution problem. These type of problems are regularly encountered in the
field of signal theory and image processing and various algorithms have been
successfully developed to address the ill-conditioned deconvolution problem
\cite{Jansson2011deconvolution}.

In this work we are using a deconvolution algorithm that is based on Bayesian inference, which we discuss in detail in Appendix 3. In Fig.~\ref{fig:SelfEnergy}, we show the lattice self-energy for a 32-site cluster by means of this methods. We can clearly observe  that the cluster and coarse-grained lattice self-energy coincide very well.

\begin{figure} \begin{center}
\includegraphics[width=0.5\textwidth]{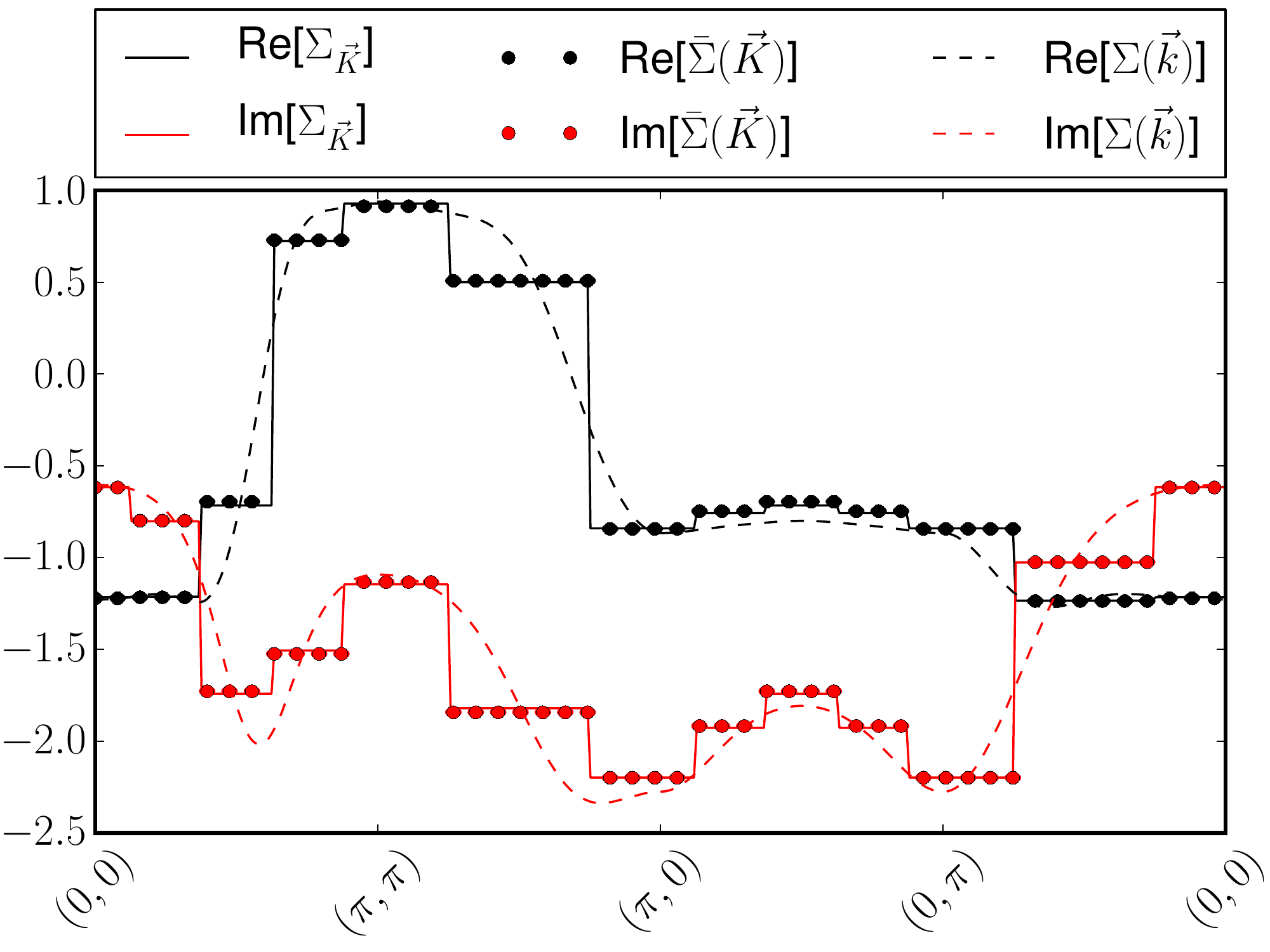} \end{center}
\caption{\label{fig:SelfEnergy} Comparison between the lattice self-energy
$\Sigma(\vec{k}, \pi\,T)$, the cluster self-energy $\Sigma_{\vec{K}}(\pi T)$
and the coarse-grained lattice self-energy at the cluster-momenta 
$\bar{\Sigma}_{\vec{K}}(\pi T) \equiv \bar{\Sigma}(\vec{K}, \pi T)$ 
for a 32-site cluster at 5\% doping and $T=0.2$.}
\end{figure}

\begin{figure*}[!] \begin{center}
\includegraphics[width=\textwidth]{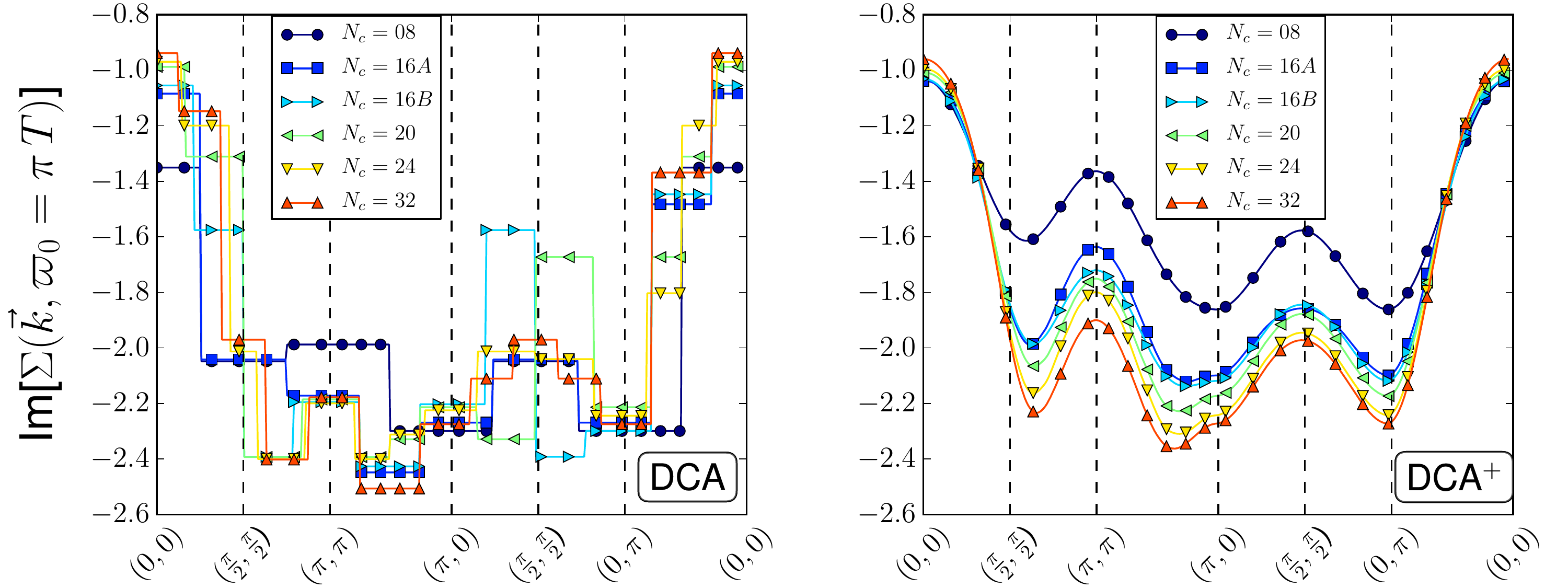}
\end{center} \caption{\label{fig:Sigma_bands} The imaginary part of the lattice
self-energy for different clusters at a temperature of $T=0.33$ with a
hole-doping of $5\%$ ($U/t=7$ and $t'/t=-0.15$).  Two key observations can be
made. The \dcaplus produces for all clusters a lattice self-energy which follows
the lattice symmetry. This is not true in the case of the DCA, which is
illustrated in the region of $(\pi, 0)$ to $(0, \pi)$ for the clusters $16B$,
$20$ and $24$. One can also observe that the \dcaplus converges monotonically.
The self-energy increases systematically with increasing cluster size as longer
range correlations are taken into account. This systematic growth of the
self-energy is harder to detect in the DCA. Therefore, we expect that the
\dcaplus will lead to a more systematic convergence of other physical
quantities, such as the pseudo gap transition temperature.} 
\end{figure*}

\section{Application:}

\subsection{Convergence of the self-energy and the pseudogap:}

One of the most distinctive features of the hole-doped cuprates is the emergence
of a pseudogap\cite{Norman:2005vr}, i.e. a partial suppression of the density of
states at the Fermi energy at the antinodal points $(\pi,0)$ and $(0,\pi)$ in
the Brillouin zone. This state appears below a temperature $T^*$, which rises
with decreasing hole doping as the Mott insulating half-filled state is
approached. The detailed relation between the pseudogap and superconductivity
remains controversial. Since superconductivity arises from the pseudogap state,
it is generally believed that understanding this unuasual phenomenon is an
important prerequisite to understanding the pairing mechanism. Recent debate has
been centered around the question of whether the pseudogap is a signature of
superconducting fluctuations above $T_c$ \cite{Emery:1995dr,Wang:2002ji} or
whether it is a competing phase\cite{Taillefer:2010gl,Gull:2012vz}.

Cluster dynamical mean field studies of the single-band Hubbard model have found
a similar pseudogap opening up at the antinodal points at low temperatures in
the low doping regime
\cite{Maier2005RMP,Macridin:2006kp,Parcollet:2004da,Kyung:2006cd,Berthod:2006fy,
Gull2010PRB}. In these calculations, the pseudogap originates from a strong
momentum-space variation of the single-particle self-energy, which, as shown in
recent DCA calculations by Gull et al.\cite{Gull2010PRB}, gives rise to a
momentum-sector-selective metal-insulator transition. The \dcaplus improves upon
the DCA algorithm in that it gives a self-energy with smooth and therefore more
physical momentum dependence, and can therefore provide new insight into this
problem. In addition, since previous studies were limited to relatively small
clusters up to 16 sites, it is important to explore whether the self-energy and
pseudogap physics is converged on such clusters.

In Fig.~\ref{fig:Sigma_bands}, we plot the imaginary part of the lattice
self-energy at the smallest Matsubara frequency $\omega_0=\pi T$ for various
clusters, computed with the DCA (left panel) and the \dcaplus (right panel). One
immediately observes the much more physical smooth momentum dependence of the
\dcaplus results versus the step-function-like nature of the DCA results for the
self-energy. At closer inspection, one notices a much more systematic
convergence of the \dcaplus results with different cluster size and geometry.
While the DCA results for ${\rm Im} \Sigma({\vec K})$ show smaller spread at a
given ${\vec K}$-point (e.g. at ${\vec K}=(\pi,0)$), their cluster dependence is
non-monotonic. In \dcaplus, in contrast,  $|{\rm Im} \Sigma({\vec K})|$
monotonically increases with cluster size -- a sensible result as longer ranged
correlations are systematically taken into account.

Another striking feature of the DCA results is the asymmetry  for clusters that
do not have the full lattice symmetry such as the 16B, 20 and 24 site clusters.
E.g., in the 16B cluster, the asymmetry around $(\pi/2,\pi/2)$ as one moves
along the line from $(\pi,0)$ to $(0,\pi)$ is apparent and the results in these
regions are significantly different from those for the symmetric 16A cluster.
This asymmetry results from the asymmetric arrangement of the two cluster
K-points closest to $(\pi/2,\pi/2)$ with respect to $(\pi/2,\pi/2)$ (see right
hand side of Fig.~\ref{fig:clusters}). This asymmetry is completely removed in
the \dcaplus.

In addition, with the exception of a small region around $(\pi,\pi)$, the
\dcaplus results for the asymmetric 16B cluster are almost identical to the
results of the fully symmetric 16A cluster. The \dcaplus algorithm restores the
full lattice symmetry in the results obtained from clusters that do not have the
full symmetry and thus makes studies on these clusters much more useful. This,
combined with the improved convergence as a function of cluster size allows for
much more systematic and precise extrapolations to the exact infinite cluster
size.

To further illustrate this point, we now turn to a study of the temperature
$T^*$ below which the pseudogap starts to form. Here, we define $T^*$ as the
maximum in the temperature dependence of the bulk ($q=0$) magnetic
(particle-hole, spin $S=1$) susceptibility $\chi_{ph}(q=0,T)$. The downturn in
this quantity below $T^*$ with decreasing temperature signals the suppression of
low-energy spin excitations, which is also observed in experiments to accompany
the opening of the pseudogap in the single-particle spectral weight. In the DCA
and \dcaplus algorithms, $\chi_{ph}$ is computed from the single and
two-particle Greens-function $G^{II}_{ph}$ obtained from the cluster-solver.
Using the notation $K = (\vec{K}, \varpi)$, the bare two-particle
Greens-function $G^{II}_{0,ph}$ is constructed from a pair of interacting
cluster Greens functions (for ${\vec q}=0$)

\begin{align} G^{II}_{0,ph}(K) &=  G(K)\:G(K) \: , \nonumber \end{align}

\noindent while the fully renormalized two-particle Green's function
$G^{II}_{ph}$ is computed as

\begin{align} &G^{II}_{ph}(K, K') = \Bigg( \prod_{l=1}^4 \int_0^{\beta} d\tau_l
\Bigg) e^{i\:\varpi_1\:(\tau_1-\tau_2)} e^{i\:\varpi_2\:(\tau_3-\tau_4)}
\nonumber \\ &\times  \sum_{\sigma, \sigma'=\pm} \langle
c^{\dagger}_{\sigma}(\vec{K}, \tau_1)\:c_{\sigma}(\vec{K},
\tau_2)c^{\dagger}_{\sigma'}(\vec{K}', \tau_3)\:c_{\sigma'}(\vec{K}',
\tau_4)\rangle. \nonumber \end{align}

The irreducible cluster vertex function $\Gamma_{ph}({\vec Q}=0,{\vec K},{\vec
K}')$ is then obtained by inverting the Bethe-Salpeter equation on the cluster

\begin{align}\label{eq:gammaph} \Gamma_{ph} &= \Big[G^{II}_{0,ph}\Big]^{-1} -
\Big[G^{II}_{ph}\Big]^{-1}, \end{align} where we used a matrix notation in in
the cluster momenta $\vec{K}$ and $\vec{K}'$. The uniform lattice spin
susceptibility $\chi_{ph}(q=0)$ is then calculated from

\begin{align} \chi_{ph} &= \sum_{K_1, K_2} \chi^0 \: [ \mathbb{1}  -
\Gamma\:\chi^0]^{-1} \nonumber. \end{align} Here, $\chi^0$ is the coarse-grained
bare susceptibility of the lattice, \begin{align} \chi^0(K) &= \:\int d\vec{k}
\: \phi_{K}(\vec{k})\: G(\vec{k})G(\vec{k})\: \nonumber \end{align}

\begin{figure}[!] \begin{center}
\includegraphics[width=0.5\textwidth]{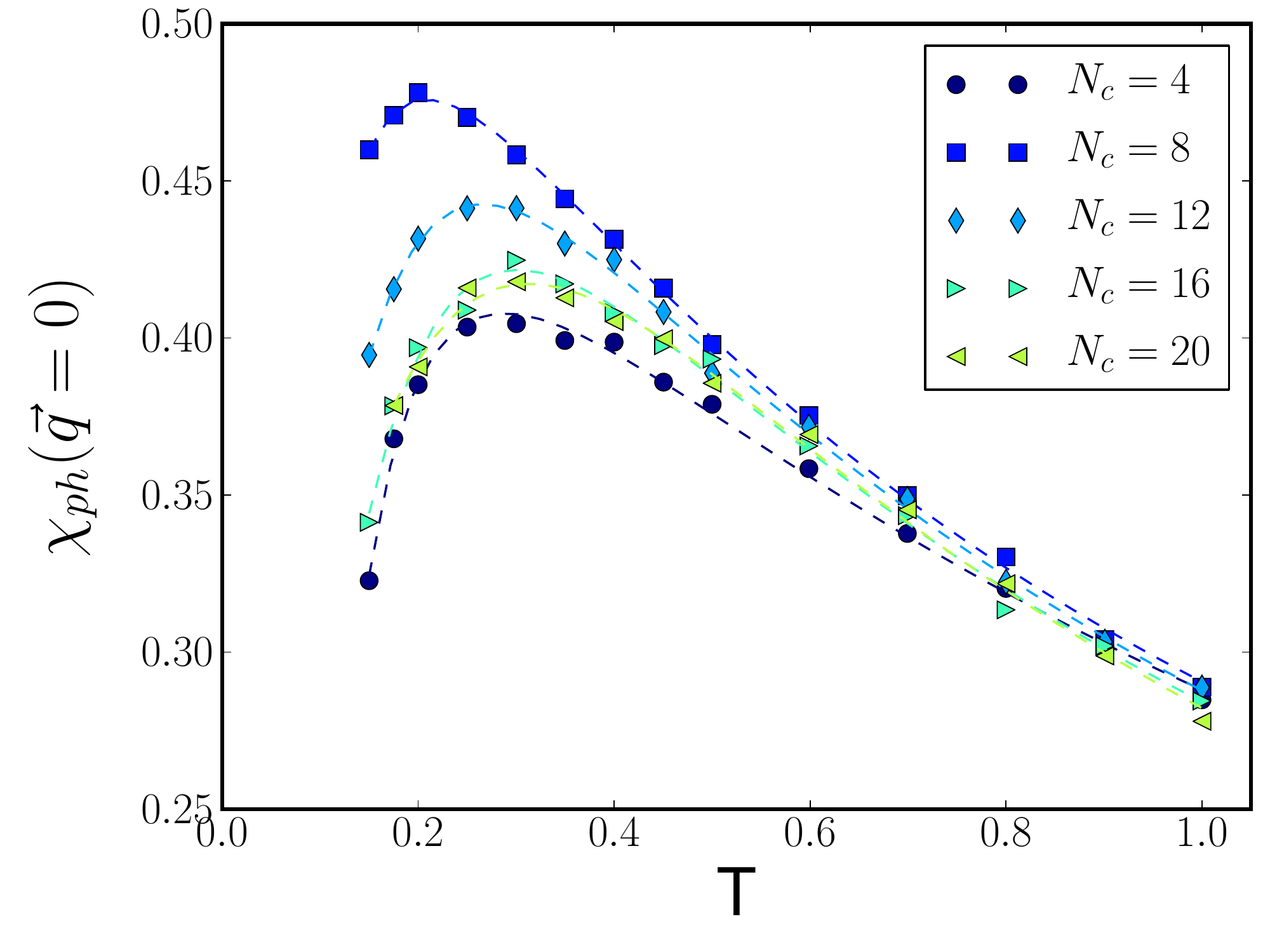}
\end{center} \caption{\label{fig:chi_DCA} Uniform spin $\chi_{ph}$
susceptibilities vs temperature for different cluster computed in the DCA at $5$
percent doping ($U/t=7$ and $t'/t=-0.15$).} \end{figure}

\begin{figure}[!] \begin{center}
\includegraphics[width=0.5\textwidth]{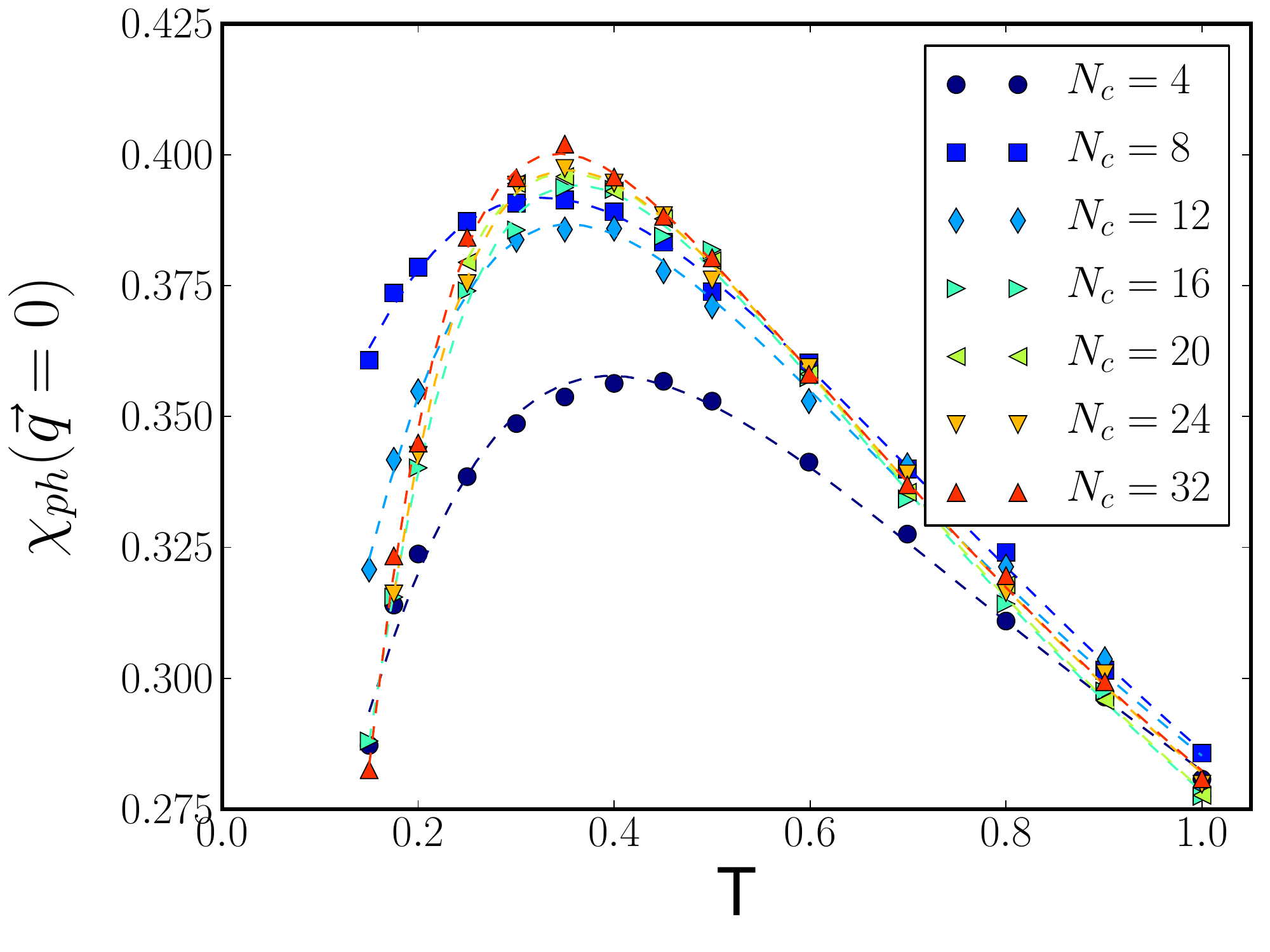}
\end{center} \caption{\label{fig:chi_DCA+} Uniform spin $\chi_{ph}$
susceptibilities vs temperature for different cluster computed in the \dcaplus
at $5$ percent doping ($U/t=7$ and $t'/t=-0.15$).} \end{figure}

\begin{figure}[!] \begin{center}
\includegraphics[width=0.5\textwidth]{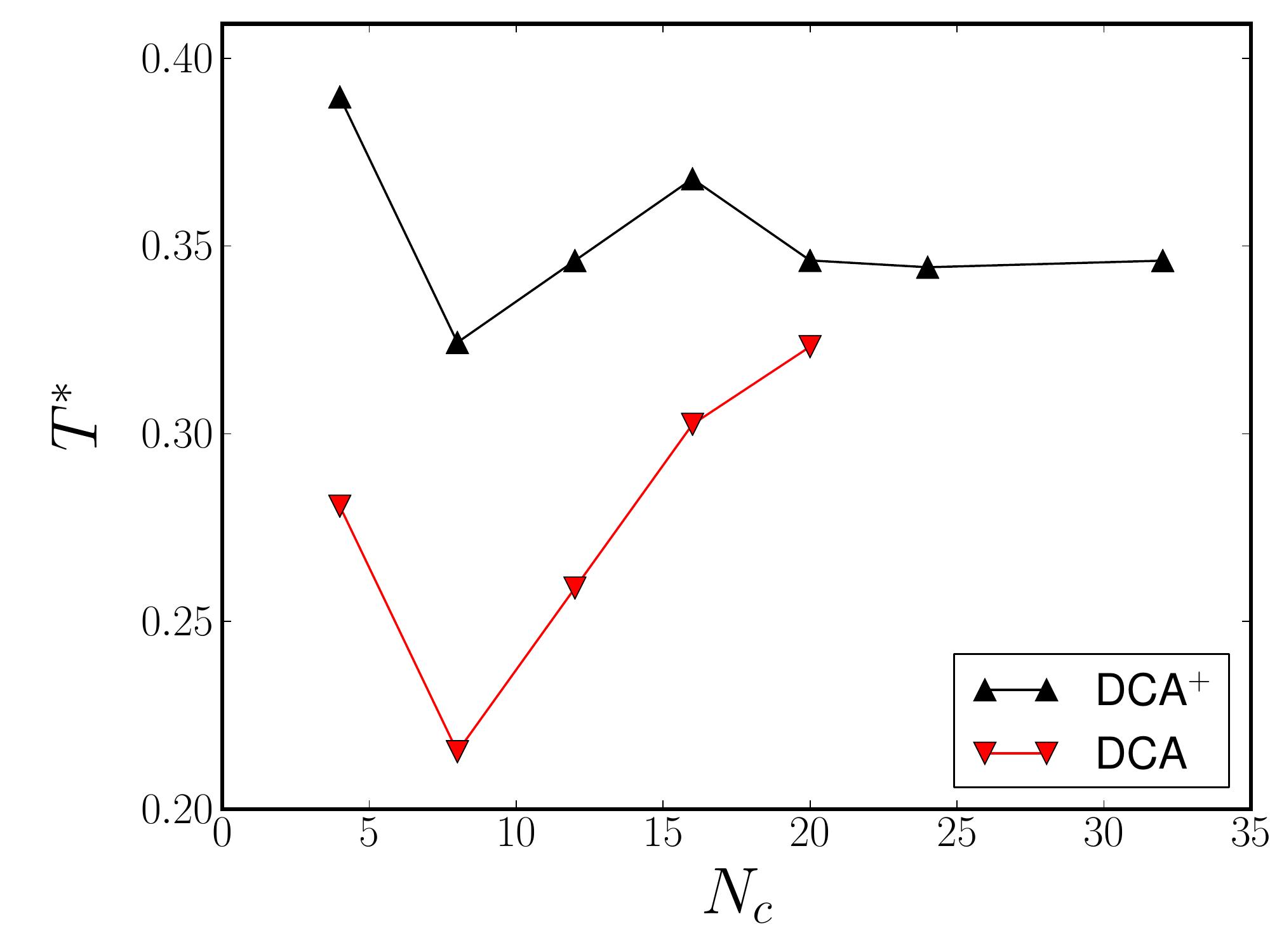} \end{center}
\caption{\label{fig:Tstar} $T^*$ versus clustersize computed in the DCA and
\dcaplus at $5$ percent doping ($U/t=7$ and $t'/t=-0.15$).} \end{figure}

This procedure to compute the uniform lattice spin susceptibility
$\chi_{ph}({\vec q}=0)$ is the same in the \dcaplus as in the
DCA\cite{Jarrell2001PRB}. The quantities that enter these equations, however,
are different between both approaches. In the \dcaplus, for thermodynamic
consistency, one should apply the same interpolation procedure to the vertex
function $\Gamma_{ph}(K,K')$ as is done for the self-energy. Here however, for
the sake of simplicity and in order to focus on the effects of the self-energy,
we keep the piecewise constant dependence of $\Gamma_{ph}(K,K')$ that is
naturally obtained from its extraction from the cluster quantities in
Eq.~(\ref{eq:gammaph}) as in the DCA. In the $S=1$ particle-hole channel, where
the leading correlations are antiferromagnetic and have only weak internal
${\vec K}$-dependence\cite{Bulut1993PRB}, we expect this to be a good
approximation.

In Fig.~\ref{fig:chi_DCA}, we show results for $\chi_{ph}({\vec q}=0)$ obtained
with the DCA for different clusters. One observes a strong cluster size
dependence and the results are not converged even for the largest cluster that
can still be simulated before the fermonic sign problem begins to make the QMC
sampling exponentially difficult. The corresponding \dcaplus results are
displayed in Fig.~\ref{fig:chi_DCA+}. Here, convergence is reached much sooner.
The location of the maximum in temperature dependence, $T^*$, is essentially
independent of the cluster for $N_c \geq 8$ (see Fig.~\ref{fig:Tstar}). As
discussed previously, this directly results from the improved convergence of the
self-energy in the \dcaplus. From these results, once the effects of cluster
geometry are removed in the \dcaplus, it becomes clear that the underlying
correlations that lead to the pseudogap formation are short-ranged and well
contained in clusters of size 8.

\subsection{Improved fermionic sign-problem}

The rapidly increasing capability of computers in conjunction with the growing
sophistication and efficiency of quantum Monte Carlo solvers has pushed the
limits of simulations to larger cluster sizes and interaction strengths, as well
as lower temperatures. As a result, the only serious barrier for quantum Monte
Carlo calculations at low temperatures and away from certain parameter regimes
(such as half-filling in the single-band Hubbard model) that remains is the
fermionic sign problem\cite{Troyer:2005ui}, which leads to an exponentially
growing statistical error with increasing system size and interaction strength,
and decreasing temperature.

The sign problem has posed an insurmountable challenge to quantum Monte Carlo
calculations of fermionic systems, especially for simulations of finite size
systems, and remains a problem in the DCA approach. The DCA, however, was shown
to have a less severe sign problem than finite size calculations
\cite{Jarrell2001PRB}, which, in the absence of a rigorous mathematical
justification, was attributed to the action of the mean-field host on the
cluster. This has enabled simulations of larger clusters at lower temperatures
than those accessible with finite size simulations and thus has opened new
possibilities for gaining insight into low temperature phenomena in correlated
systems.

The \dcaplus approach is different from the DCA in that it generates a more
physical self-energy with smooth momentum dependence, and the correlations
described by this self-energy are therefore shorter-ranged than those in the
DCA. Hence, it is therefore not unreasonable to expect a difference in the
severity of the sign problem between \dcaplus and DCA.

\begin{figure}[!] \begin{center}
\includegraphics[width=0.5\textwidth]{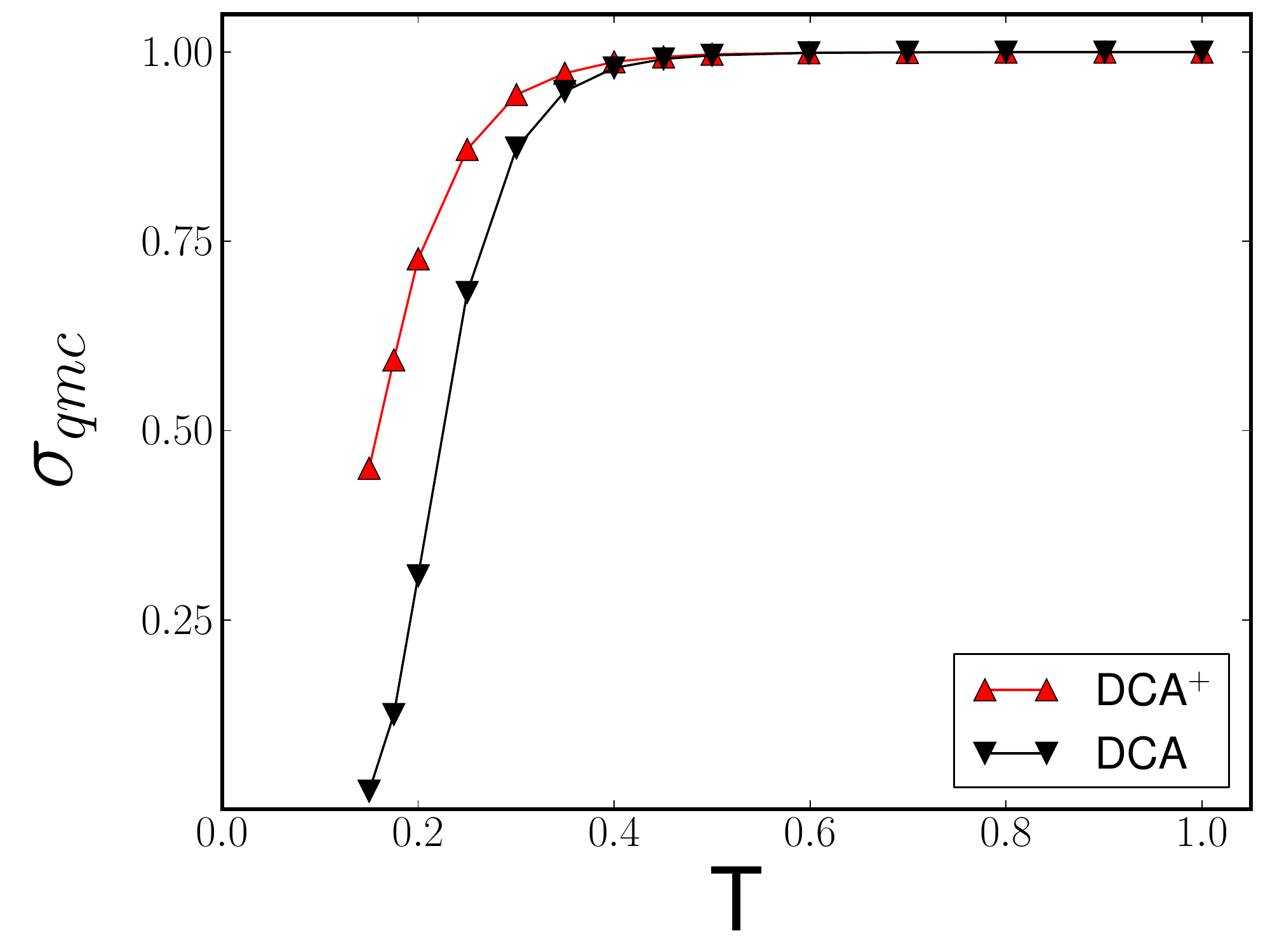} \end{center}
\caption{\label{fig:fermionic_sign} Temperature dependence of the average
fermionic sign for $N_c=32$ at $5$ percent doping ($U/t=7$ and $t'/t=-0.15$).}
\end{figure}

In Fig.~\ref{fig:fermionic_sign} we compare the fermionic sign $\sigma_{qmc}$
between the DCA and the \dcaplus for a 32-site cluster and $U=7t$ for a doping
of 5\%. At low temperatures, the average sign in the \dcaplus simulation is
significantly larger than that of the DCA simulation. As indicated above, we
attribute this improvement to the smooth momentum dependence of the \dcaplus
self-energy as compared to the step function dependence of the DCA self-energy.
From Fourier analysis, one knows that the smoothness of a function is related to
the rate of decay of its Fourier coefficients\cite{katznelson2004}. More
precisely, if a function $f$ is $p$ times differentiable, then its Fourier
components $f_n$ will decay at least at a rate  of $~1/n^{p+1}$

\begin{align} f \in C^p \quad \rightarrow \quad \vert f_n \vert \leq \frac{\vert
f^{(p)} \vert_1}{n^{p+1}}. \end{align}

Since the \dcaplus self-energy has smooth momentum dependence and not the step
discontinuities of the DCA, its Fourier-transform to real space is
shorter-ranged than that of the DCA and the correlations it describes are
shorter-ranged. We believe that it is this removal of unphysical long-range
correlations, which reduces the sign problem in the \dcaplus. In any case, with
this significant reduction in the severity of the sign problem, it is possible
to study the physics of fermionic systems in even larger clusters and at lower
temperatures than accessible with the DCA.

\section{Summary and Conclusions}

In this paper, we have presented the theoretical framework as well as a
practical implementation of the \dcaplus algorithm. It is an extension to the
DCA without the jump discontinuities inherent  in the standard DCA algorithm
that computes a continuous lattice self-energy in a self-consistent way. This
improvement is based on two fundamental differences to the DCA. First, an
explicit distinction is made between the lattice and the cluster self-energy.
Second, a continuous lattice self-energy is determined in a way so that its
coarse-grained value $\bar{\Sigma}_{\vec{K}}$ is equal to the cluster
self-energy $\Sigma_{\vec{K}}$ obtained from the cluster-solver.  This
constraint makes the \dcaplus algorithm fundamentally different from previous
attempts\cite{Hettler2000PRB, Okamoto2003PRB} to include a continuous
self-energy into the DCA self-consistency loop that lead to an acausal and thus
a non-physical self-energy. 
during the coarse-graining of the Greens-function but itself has not been
coarse-grained.

The new coarse-graining rules in the cluster-mapping of \dcaplus require us to
reconsider the lattice-mapping in the algorithm. As a matter of fact, we have
shown that a continuous lattice self-energy $\Sigma(\vec{k})$ can only be
inferred from the discrete cluster self-energy $\Sigma_{\vec{K}}$ if the DCA
assumption of smoothness of the lattice self-energy is satisfied. This has been
discussed in the paper using the properties of the projection operator $P_{i,j}$
 that is associated with the coarse graining operation in
Eq.~(\ref{cluster_mapping_2}). The transformation of the cluster self-energy
into the lattice self-energy amounts to inversion of the projection operators
$P_{i,j}$. Since this is a singular operator, the lattice mapping is only
well-defined as long as the cluster self-energy converges on the image-space of
the operator, which is spanned by the eigenvectors with non-zero eigenvalue. In
practice the image-space is the space spanned by eigenvectors with an eigenvalue
larger than a given parameter $\epsilon$. The convergence behavior of the
\dcaplus algorithm is determined by two essential properties of the projectors
$P_{i,j}$: (1) the dimension of the image-space increases with cluster size,
which is consistent with the intuitive notion that larger cluster can support
finer features of the self-energy; (2) the delocalization of each eigenvector
$\langle r^2 \rangle$ and the magnitude of its corresponding eigenvalue are anti
correlated. Consequently, for large cutoff parameter $\epsilon$ a more localized
cluster self-energy is needed in order to have a controlled lattice mapping.
Self-consistency in the \dcaplus can only be reached if the cluster self-energy
is localized enough to converge on the image space of the projection operator.
If convergence is not reached, the image space of the projector and thus the
cluster size will have to be increased. Convergence thus provides a useful
measure for the quality of a \dcaplus calculation with a given cluster.

Straightforward inversion of the projection operator would be numerically
unstable, since the projection operator is a near singular matrix. Thus, in the
implementation of the lattice mapping in the \dcaplus algorithm we have followed
a different approach, splitting it into two numerically stable steps. First, we
interpolate the cluster self-energy in a controlled way, using an injective
transformation, and next, we deconvolute this interpolated, continuous cluster
self-energy using the Richardson-Lucy algorithm. In both steps convergence
within the self-consistent loop can be monitored by an objective measure. For
the interpolation we know that the Fourier transform of
$\mathcal{T}[\Sigma_{\vec{K}}] = (\Sigma_{\vec{K}}-\imath)^{-1}$ has to converge
on the real-space impurity cluster in order to obtain an accurate interpolation.
For the deconvolution, the difference between the coarsegrained lattice
self-energy $\bar{\Sigma}_{\vec{K}}$ and the cluster self-energy
$\Sigma_{\vec{K}}$ has to be smaller than the statistical error of the
Monte-Carlo integration.

To illustrate the benefits of the \dcaplus algorithm we have investigated the
pseudogap phase in a lightly hole-doped two-dimensional Hubbard model. Like with
the DCA, the \dcaplus based calculations give a self-energy that has strong
momentum dependence. However, we find that the \dcaplus has a much reduced
fermionic sign problem and thus we can investigate the pseudogap phase on larger
clusters and in more details than in the standard DCA. In the \dcaplus the
self-energy is continuous in momentum space and thus more physical, and it
converges monotonically and much more systematically with  cluster size than in
the DCA. A similarly improved convergence behavior in the \dcaplus is found for
the pseudogap temperature $T^*$ below which the bulk lattice susceptibility
decreases with decreasing temperature. In the DCA, we find that $T^*$ has a
strong cluster dependence and converges only for the largest possible cluster
sizes. In the case of the \dcaplus, we observe a much faster convergence of
$T^*$, which is a direct consequence of the improved convergence of the
self-energy in the \dcaplus. From the convergence property of $T^*$, we can
conclude that the correlations responsible of the pseudogap formation must be
short ranged and well contained in a cluster size of eight sites. This improved
convergence in the \dcaplus raises the hope to do precise extrapolations to the
exact infinite cluster size limit in future calculations of other properties.

\begin{acknowledgments}

This research was carried out with resources of the Swiss National
Supercomputing Center (CSCS), Oak Ridge Leadership Computing Facility (OLCF),
and the Center for Nanophase Materials Sciences (CNMS). OLCF and CNMS are
located at Oak Ridge National Laboratory and supported respectively  by the
Office of Science under Contract DE-AC05-00OR22725 and by the Scientific User
Facilities Division, Office of Basic Energy Sciences, of the Department of
Energy. CSCS is an autonomous unit of ETH Zurich.

\end{acknowledgments}

\section{Appendix 1: Analysis of the projection operator $P_{i,j}$ and its connection to
the locality of $\Sigma(\vec{k})$.}

In this Appendix, we give the reader an intuitive
understanding of the projection operator $P_{i,j}$ that plays a central role in the cluster mapping procedure of the \dcaplus algorithm. 
We show that its inverse exists if the DCA locality assumption is satisfied 
for the lattice self-energy. Furthermore, we
discuss how the projection operator $P_{i,j}$ is influenced by the choice of the
cluster.

To this end, we expand the lattice self-energy in terms of cubic Hermite
splines\cite{Keys1981}. These functions form a basis for cubic splines and obey
a convolution property. The lattice self-energy can therefore be written as sum
over a very fine mesh $\{\vec{k}_i\}$ in momentum space.

\begin{equation}\label{herm_DCA} \Sigma(\vec{k}) = \sum_{\vec{k}_i}
\sigma_{\vec{k}_i} \: \mathcal{H}(\vec{k}-\vec{k}_i) \quad \mbox{with} \quad
\Sigma(\vec{k}_i) = \sigma_{\vec{k}_i} \end{equation}

\begin{figure}[t] \begin{center}
\includegraphics[width=0.5\textwidth]{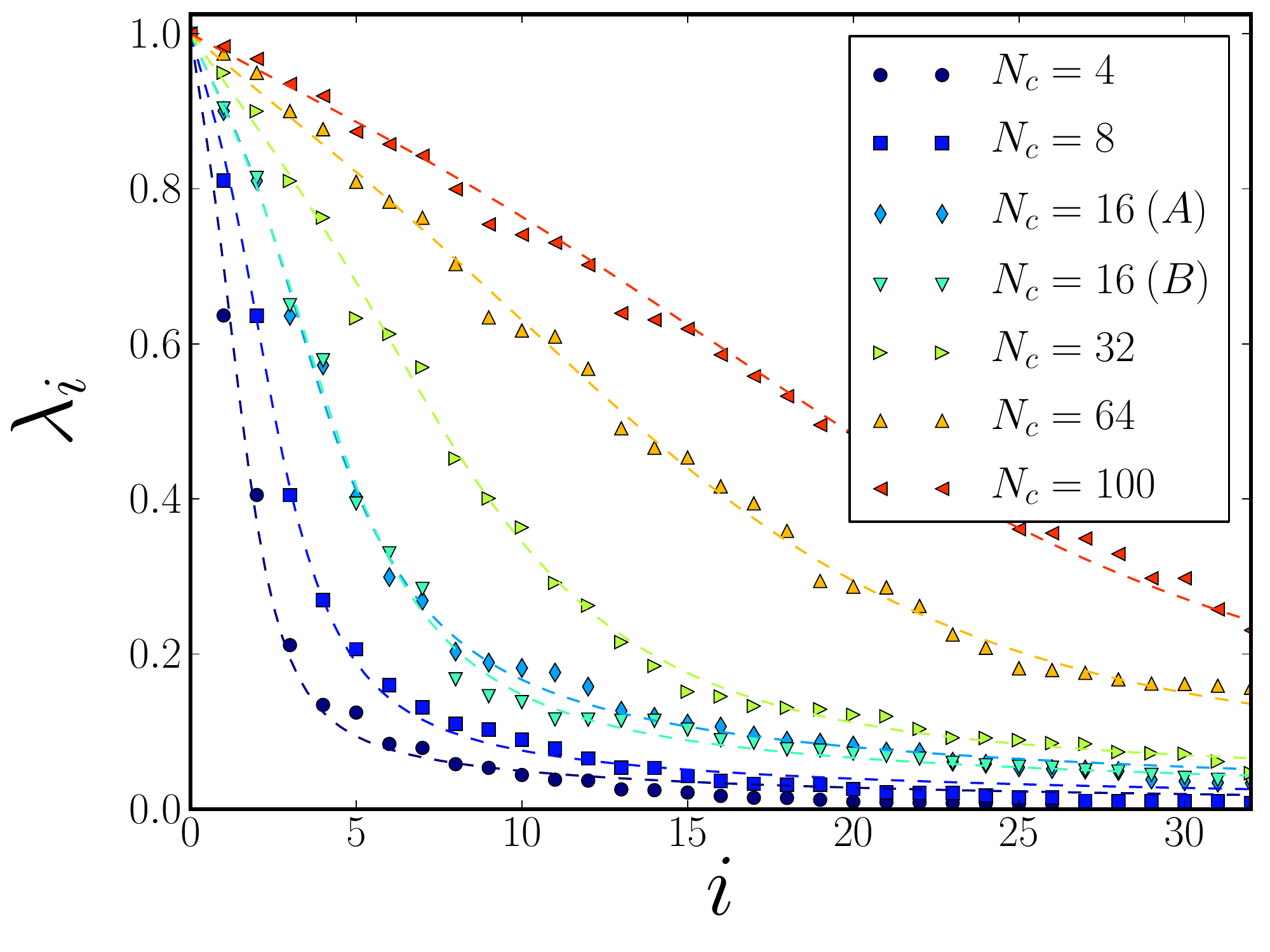} \end{center}
\caption{\label{fig:eigenvalues} The leading eigenvalues of various clusters on
a fine mesh of 512 points. We can clearly observe a strong decay of the leading
eigenvalues for small clusters, which becomes weaker with increasing the
cluster-size. This observation explains the intuitive notion that large clusters
can describe finer features in the self-energy, since the image-space of larger
clusters contains more eigenvectors.} \end{figure}

It has to be stressed that choosing Hermite splines as a basis will not
influence the conclusions we obtain here and thus does not reduce the generality
of our arguments. It just simplifies the discussion, since the expansion index
$i$ can now  be identified with a lattice momentum $\vec{k}_i$ in the fine
lattice mesh and the expansion coefficient $\sigma_{i}$ with the lattice
self-energy at that lattice momentum $\vec{k}_i$. Next, we generalize the
cluster-mapping in Eq.~(\ref{cluster_mapping_2}), by replacing the cluster
momentum points $\{\vec{K}_i\}$ by the fine lattice $\{\vec{k}_i\}$. The
coarse-graining then becomes a convolution of the lattice self-energy with the
patches and we obtain

\begin{equation}\label{proj_DCA} \bar{\Sigma}_{\vec{k}_i} = \sum_{j}
\sigma_{\vec{k}_j} \:  \underbrace{\int d\vec{k}\:
\phi_{0}(\vec{k}-\vec{k}_i)\:\mathcal{H}(\vec{k}-\vec{k}_j)}_{=P_{\vec{k}_i,
\vec{k}_j}}. \end{equation}

The projection-matrix $P_{\vec{k}_i, \vec{k}_j}$ has now become a symmetric,
square matrix. The latter allows us to do a spectral decomposition of
$P_{\vec{k}_i, \vec{k}_j}$ into its eigenspace. If we represent its eigenvalues
by $\lambda$ and its corresponding eigenvector by $e_{\lambda}$, we obtain

\begin{equation}\label{spec_DCA} {\bar \Sigma}_{\vec{k}_i} = \sum_{j}
\sigma_{\vec{k}_j} \:  \sum_{\lambda} \lambda\: e_{\lambda}(\vec{k}_i) \times
e_{\lambda}^{T}(\vec{k}_j) \end{equation}

In terms of the eigenspace of the projection-operator, the cluster- and
lattice-mapping can now be written as

\begin{align}\label{DCAplus_mappings} &\mbox{cluster-mapping:}\quad
\bar{\Sigma}_{\vec{k}_i} = \sum_{\lambda} \lambda \: \langle
{\sigma}_{\vec{k}_j},  e_{\lambda}(\vec{k}_j) \rangle \: e_{\lambda}(\vec{k}_i)
\:   \nonumber\\ &\mbox{lattice-mapping:}\quad\sigma_{\vec{k}_i} =
\sum_{\lambda}  \lambda^{-1} \: \langle \bar{\Sigma}_{\vec{k}_j},
e_{\lambda}(\vec{k}_j) \rangle \: e_{\lambda}(\vec{k}_i) \: \end{align}

\begin{figure}[t] \begin{center}
\includegraphics[width=0.5\textwidth]{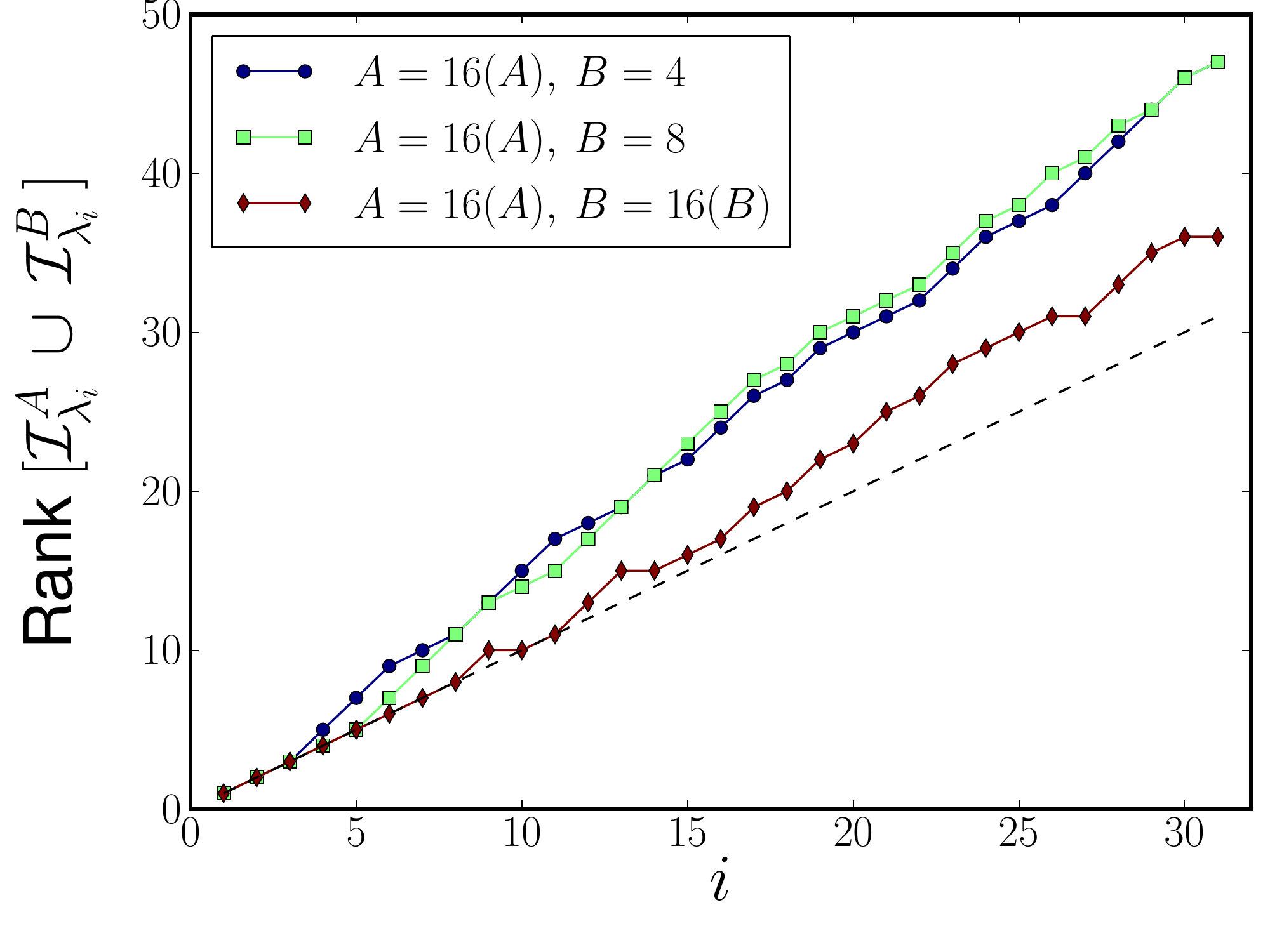} \end{center}
\caption{\label{fig:eigenvectors_diff} The dimension of the union image space
$\mathcal{I}^A_{\lambda_i} \cup \mathcal{I}^B_{\lambda_i}$ for two different
clusters A and B versus the eigenvalue index $i$. Since the rank of
$\mathcal{I}^A_{\lambda_i}$ and $\mathcal{I}^B_{\lambda_i}$ both equal $i$, any
deviation of the rank for the space $\mathcal{I}^A_{\lambda_i} \cup
\mathcal{I}^B_{\lambda_i}$ from $i$ indicates that the projection operators of
clusters $A$ and $B$ span different image spaces. One can clearly observe that
the differentiation of the $16A$ site cluster eigenspace with smaller clusters
occurs faster.} \end{figure}

Here, the inner-product $\langle \vec{a} , \vec{b} \rangle$ is represented by a
simple dot-product between the two vectors  $\vec{a}$ and $\vec{b}$. From
Eqs.~(\ref{DCAplus_mappings}), it is clear that the spectrum $\{\lambda\}$ of
the projection-operator $P_{ij}$ plays a central role in the cluster- and
lattice mapping. In Fig.~\ref{fig:eigenvalues}, we show the leading eigenvalues
(i.e. having the largest absolute value) of $P_{i,j}$ for various clusters. One
can clearly observe that all eigenvalues are smaller or equal than one and decay
rapidly for small clusters ($N_c \leq 8$) and slowly for large clusters ($N_c
\geq 32$). This can be easily understood from the form-factor of the patches.
The latter are very similar to box-car filters, which are one of the most common
low-pass filters used in the field of signal processing. Since the
coarse-graining of the lattice self-energy in Eq.~(\ref{proj_DCA})  can be
rewritten as a convolution with the patches, the projection operator $P_{i,j}$
will in fact reduce all the Fourier components during the convolution, insuring
that the $L_2$-norm of any function in the eigenspace never grows. Consequently,
this is also true for all eigenvectors, which leads us to conclude that the
eigenvalues have to be less or equal to 1.

With the spectral decomposition of the projection matrix 
we can split the representation space of the continuous lattice self-energy into the image-space
$\mathcal{I}$ and the kernel-space $\mathcal{K}$ of the projection operator
$P_{i,j}$. Since our projection-operator does not follow the strict mathematical
definition of a projection operator\footnote{A projection operator should
satisfy the relationship $P^2=P$. The eigenvalues of such an operation can only
be $0$ and $1$. }, we define the image $\mathcal{I}_{\epsilon}$ as the space
spanned by the eigenvectors that have an eigenvalue larger than $\epsilon$.
Here, $\epsilon$ is a small, positive cut-off parameter. The kernel
$\mathcal{K}_{\epsilon}$ contains the remainder of the space, and is thus
spanned by the eigenvectors with an eigenvalue smaller than $\epsilon$. 
Due to the inversion of the eigenvalue in Eq.~(\ref{DCAplus_mappings}), the
lattice-mapping is only well-defined on the image-space
$\mathcal{I}_{\epsilon}$. This brings us to the first important observation. In
order to do a self-consistent \dcaplus calculation, the coarse-grained lattice
self-energy should always be entirely defined on the image-space
$\mathcal{I}_{\epsilon}$ of our projection operator. Otherwise, there exists no
well-defined transformation that maps the cluster self-energy back into the
lattice self-energy, which in turn breaks the \dcaplus self-consistency loop.
Notice that this requirement holds trivially in the case of the traditional DCA,
since in that case the projection matrix is simply the identity-matrix of size
$N_c$, and all eigenvalues are equal to one.

Eq.~(\ref{spec_DCA}) can also explain how the geometry of the patches will
influence the results obtained with the \dcaplus. In
Fig.~\ref{fig:eigenvectors_diff}, we plot the union space of the image spaces
$\mathcal{I}_{\lambda_i}^A$ and $\mathcal{I}_{\lambda_i}^B$ versus eigenvalue
index $i$ for different clusters. The plot shows very clearly that the first
leading eigenvectors are equal to each other, and gradually diverge as
eigenvectors with smaller eigenvalues are added. This brings us to the second
observation. If one wants to carry out a DCA-calculation with results that are
independent of cluster shape, the cluster self-energy has to be representable on
the intersection of the image-spaces $\mathcal{I}_{\epsilon}$ of both clusters.

\begin{figure}[t] \begin{center}
\includegraphics[width=0.5\textwidth]{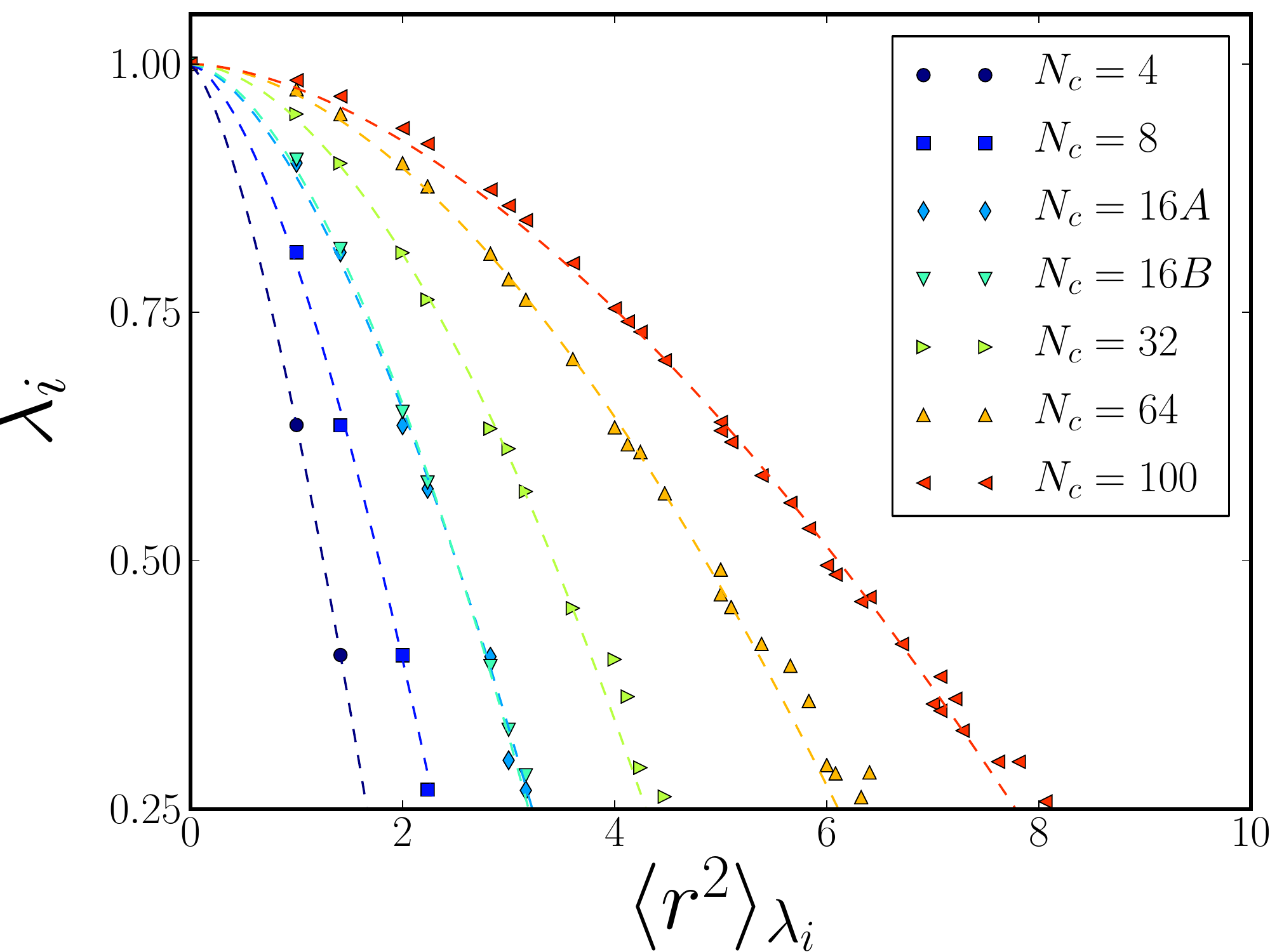} \end{center}
\caption{\label{fig:r_versus_lambda}  The correlation between the magnitude of
the leading eigenvalue and the delocalization of the its corresponding
eigenvector for various clusters. } \end{figure}

So far, we have only discussed and introduced strict geometrical criteria on the
lattice and cluster self-energy, that indicate when a \dcaplus cluster
calculation is feasible. In order to link geometrical criteria to physics, we
show in Fig.~\ref{fig:r_versus_lambda} the delocalization of the leading
eigenvectors $\langle r^2 \rangle$. Formally, we define the delocalization as

\begin{equation}\label{deloc_DCA} \langle r^2 \rangle_{\lambda} =
\sqrt{\frac{\sum_{\vec{r}} e_{\lambda}^{T}(\vec{r}) \: r^2 \: 
e_{\lambda}(\vec{r}) }{\sum_{\vec{r}}
e_{\lambda}^{T}(\vec{r})\:e_{\lambda}(\vec{r}) }}. \end{equation}

At close inspection, we can see a clear correlation between the absolute value
of the leading eigenvalues $\lambda$ and the delocalization of its corresponding
eigenvector for all cluster sizes. This correlation shows that the space
$\mathcal{I}_{\epsilon}$ is actually spanned by the eigenvectors with a small
delocalization. As a result, satisfying the geometric criteria to do a
self-consistent \dcaplus calculation is essentially equivalent to satisfying the
DCA-assumption of locality for the lattice self-energy. Another important
conclusion that can be drawn from Fig.~\ref{fig:r_versus_lambda} is that the
number of vectors that span the space $\mathcal{I}_{\epsilon=0.25}$ becomes
larger with increasing cluster size. This correlation reflects the intuitive
notion in the DCA that larger clusters can describe finer features of the
lattice self-energy.

\section{Appendix 2: A Mathematical basis for the interpolation procedure.}

In this appendix, we want to demonstrate that the interpolation procedure presented in this paper is independent of the proposed transformation function $\mathcal{T}$, as long as the latter is analytical and injective. To accomplish this goal, we construct a function $g(k)$, defined by the transformed real-space Fourier components of an arbitrary function $\mathcal{F}$ that fall within a cut-off parameter $R_c$. The goal is now to show that $g(k)$ can approximate the function $\mathcal{F}$ with arbitrary precision, given a big enough cut-off parameter $R_c$. In other words, point-wise convergence of $g(k)$ towards $\mathcal{F}$ is thus guaranteed. The rate of convergence will depend crucially on the rate of convergence of $\mathcal{T}[\mathcal{F}]_R$ versus the radius $\vert R \vert$.
\newline

\textbf{Pointwise convergence:} \textit{Consider a function $\mathcal{F}$ in the Brillouin zone $\mathbb{B}$ and an
injective, continuous transformation $\mathcal{T}$, such that the Fourier
components $\mathcal{T}[\mathcal{F}]_R$  fullfill,}

\begin{align}\label{prop1} \forall \epsilon > 0,  \exists R_c \in \mathbb{R}:
\sum_{|R| \geq R_c}  \Big \vert \mathcal{T}[F]_R\Big \vert  \leq \epsilon
\nonumber \\ \mbox{with} \quad  \mathcal{T}[F]_R = \int_{\mathbb{B}} d\vec{k}
e^{-i k R} \mathcal{T}[F(\vec{k})] \end{align}

\textit{then,}

\begin{align} \forall \vec{k} \in \mathbb{B}, \, \forall \epsilon > 0,\, \exists
R_c \in \mathbb{R} :  |g(k)-\mathcal{F}(k)| < \epsilon \nonumber \\ \mbox{with}
\quad  g(k) = \mathcal{T}^{-1} \Big[ \sum_{R<R_c} \exp(\imath R k)
\mathcal{T}[F]_R \Big] \end{align}


Choose a positive small number $\epsilon$. Since $\mathcal{T}$ is a continuous
and invertible function, we know that the $\mathcal{T}^{-1}$ is also continuous.
Hence, by definition of the this continuity,  there exists a $\delta \in
\mathbb{R}^+_0$ for this $\epsilon$, such that

\begin{align}
  |\mathcal{T}[g(k)]- \mathcal{T}[\mathcal{F}(k)] | < \delta \rightarrow
  ||g(k)-\mathcal{F}(k)| < \epsilon. \nonumber
\end{align}

Using the property in Eq.~(\ref{prop1}), we can find a radius $R_c > 0$, such
that

\begin{align} \sum_{|R| \geq R_c}  \Big \vert \mathcal{T}[\mathcal{F}]_R\Big \vert  <
\delta. \end{align}

By the definition of $g(k)$, we have,

\begin{align}
  |\mathcal{T}[g(k)]- \mathcal{T}[\mathcal{F}(k)] | &=  \Big \vert  \sum_{R \geq
  |R_c} \exp(\imath R k) \mathcal{T}[\mathcal{F}]_R \Big \vert \nonumber   \\
  									&\leq   \sum_{R \geq R_c} \Big \vert
  									\mathcal{T}[\mathcal{F}]_R \Big \vert  \nonumber \\
  									&\leq \delta. \end{align}

\section{Appendix 3: The Richardson-Lucy algorithm}

One of the most common deconvolution algorithms is the Richardson-Lucy
algorithm\cite{Richards1972, Lucy1974} that is based on a Bayesian
inference scheme. Since the patches are strictly positive and integrate to
unity, we can interpret them as a probability distribution function.

\begin{align} \forall \vec{k}, \vec{k}': \: \phi_{\vec{0}}(\vec{k}-\vec{k}') \ge 0,
\qquad 1= \frac{N_c}{V_{BZ}} \: \int_{BZ} d\vec{k} \: \phi_{\vec{0}}(\vec{k}-\vec{k'})
 \nonumber \end{align}

As such, we can apply Bayes theorem and construct a conditional probability
$\mathcal{Q}$ for any given lattice self-energy  $\Sigma(\vec{K})$

\begin{align}\label{conditional_prob} \mathcal{Q}(\vec{k} | \vec{k}') &=
\frac{\phi_{0}(\vec{k}'-\vec{k})\: \Sigma^t_{l}(\vec{k})}{\int_{BZ} d\vec{k}''
\: \phi_{0}(\vec{k}'-\vec{k}'')\: \Sigma(\vec{k}'')}. \end{align}

\noindent We should stress at this point that conditional probability
$\mathcal{Q}$ is computed separately for the real and imaginary part of the
self-energy. The conditional probability $\mathcal{Q}(\vec{k} | \vec{K})$ is
then used to construct a new lattice self-energy $\Sigma^{'}(\vec{k})$, given a
continuous cluster self-energy $\bar{\Sigma}(\vec{k}')$,

\begin{align}\label{deconvolution_guess} \Sigma^{'}(\vec{k}) =  \int_{BZ}
d\vec{k}' \: \mathcal{Q}^t(\vec{k} | \vec{k}') \bar{\Sigma}(\vec{k}').
\end{align}

The idea of the Richardson-Lucy algorithm is now to use
Eq.~(\ref{conditional_prob}) and Eq.~(\ref{deconvolution_guess}) in an iterative
way. After plugging both equations together, we end up with a fixed point
problem

\begin{align}\label{deconvolution_RL} \Sigma(\vec{k}) \leftarrow 
\Sigma(\vec{k}) \int d\vec{k}' \:\frac{\phi_{0}(\vec{k}-\vec{k}')\: 
\bar{\Sigma}(\vec{k}')}{\int d\vec{k}'' \: \phi_{0}(\vec{k}'-\vec{k}'')\:
\Sigma(\vec{k}'')}. \end{align}

\begin{figure}[!] \begin{center}
\includegraphics[width=0.5\textwidth]{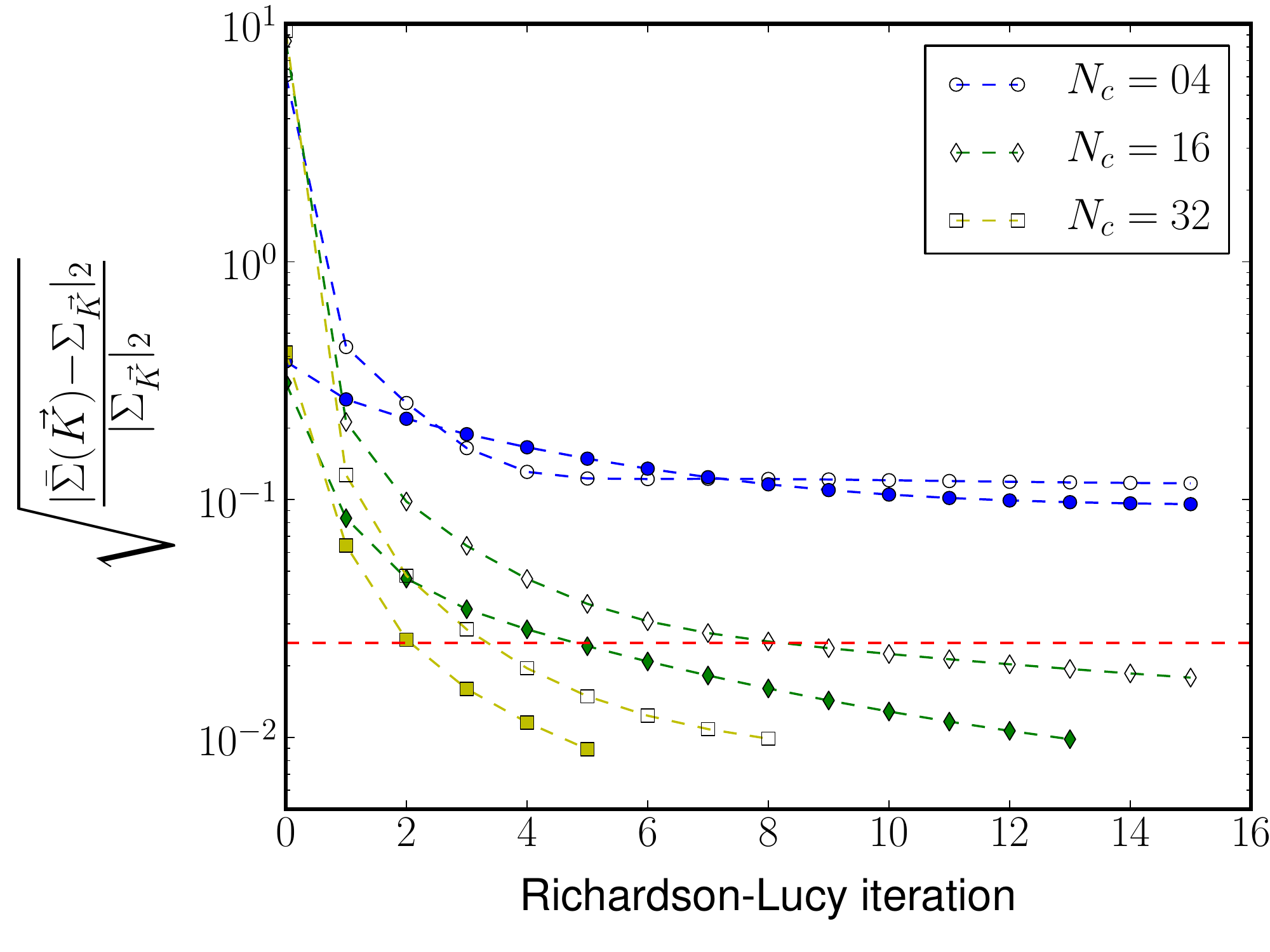} \end{center}
\caption{\label{fig:RL_error} Relative error between the cluster self-energy
$\Sigma_{\vec{K}}$ and the integrated lattice self-energy ${\bar
\Sigma}(\vec{K})$ for the real (open symbols) and imaginary (solid symbols) part
at 5\% doping and $T=0.2$.} \end{figure}

If the interpolated function $\bar{\Sigma}(\vec{k})$ is now used as our initial
guess for the lattice self-energy $\Sigma(\vec{k})$,
Eq.~(\ref{deconvolution_RL}) provides us with a simple implementation for the
lattice-mapping. In light of the \dcaplus algorithm, the Richardson-Lucy
deconvolution algorithm has many interesting properties, that make it an ideal
algorithm to be used for the deconvolution. First of all, it is a
straightforward algorithm that does not need any extra, non-physical input.
Other deconvolution algorithms, such as total variation\cite{Rudin1992,
BioucasDias2006} introduce non-physical penalty factors to insure smoothness of
the result. Secondly, the Richardson-Lucy algorithm conserves the sign of
strictly positive and negative functions. This property can be easily proven in
Eq.~(\ref{deconvolution_RL}), since $\phi_0(\vec{k})$ is strictly positive.
Hence, if the initial guess for $\Sigma(\vec{k})$ and $\bar{\Sigma}(\vec{k}')$
are both positive (negative) for all momenta $\vec{k}$, the resulting
$\Sigma(\vec{k})$ will also be positive (negative).  Therefore, if the
interpolated cluster self-energy $\bar{\Sigma}(\vec{k})$ is causal, the lattice
self-energy will also be a causal function. Third, it has been proven that the
solution of this iterative scheme converges to the maximum of the likelihood
function\cite{Lucy1974}. Hence, of all lattice self-energies that generate the
same cluster self-energy after the convolution (coarse-graining), the
Richardson-Lucy algorithm will produce the lattice self-energy that is the most
likely to reproduce the cluster self-energy.

Like all other deconvolution algorithms, the Richardson-Lucy algorithm is an
approximate algorithm, meaning that the convergence to the exact solution is not
guaranteed up to an arbitrary precision. This is not surprising, since we know
that the convolution is invertible as long as the expansion coefficients of the
cluster-self-energy in Eq.~(\ref{DCAplus_mappings}) decay faster than the
eigenvalues of the projection-operator. Consequently, the smaller the cluster,
the slower the Richardson-Lucy algorithm will converge to a solution and the
bigger the discrepancy between the coarsegrained lattice self-energy
$\bar{\Sigma}(\vec{K})$ and the cluster self-energy $\Sigma_{\vec{K}}$ obtained
from the cluster-solver. This phenomenon is illustrated in
Fig.~\ref{fig:RL_error}, where we show the relative error in the $L_2$-norm
between  $\bar{\Sigma}(\vec{K})$ and $\Sigma_{\vec{K}}$. The figure clearly
shows that the larger cluster converges faster and that the residual error
between the cluster and coarsegrained self-energy decreases with increasing
cluster-size.

\bibliography{./correct_references.bib}
\end{document}